\documentstyle[prl,aps,floats,epsf,epsfig]{revtex}

\begin{document}

\draft
%\tighten

\title{
%Non Perturbative Thermodynamics 
Avoided Phase Transitions and Glassy Dynamics
In Geometrically Frustrated Systems and Non-Abelian
Theories}

%\title{Thermodynamics Incurred by Non-Abelian Fields And 
%Viable Consequences For Supercooled Liquids}

%\title{Avoided Critical Points In Non-Abelian Actions And Supercooled Liquids}

\author{Zohar Nussinov*} 
\address{Institute Lorentz for Theoretical Physics, Leiden University\\
P.O.B. 9506, 2300 RA Leiden, The Netherlands}
\date{\today ; E-mail:zohar@lorentz.leidenuniv.nl}

\twocolumn[

\widetext
\begin{@twocolumnfalse}

\maketitle

\begin{abstract}
We demonstrate that the application of any external
uniform non-abelian gauge background, no matter
how small, leads to a greatly enhanced degeneracy. 
This degeneracy is so large that even a non-abelian 
background field of infinitesimal strength 
leads to a shocking change in the 
thermodynamics. The critical temperature 
might be discontinuously depressed and an ``avoided
critical point'' will emerge. We 
focus on how this arises in models 
previously employed to describe the microscopics 
of metallic glasses and correctly predicted the structure
factor peaks. Some of the best fits, to date, 
to the dynamics of supercooled liquids 
were inspired by such notions for 
which we now provide a suggestive 
microscopic basis. We generalize the
Mermin-Wagner inequality to high dimensions
and discuss how extensive configurational entropy 
may be computed, by replica calculations, 
for a multitude of glass models
(including non-Abelian gauge
backgrounds). This extensive
configurational entropy 
then allows a possible derivation of 
Vogel-Fulcher dynamics. 
We  fortify earlier 
ideas suggesting avoided 
critical dynamics.

\end{abstract}

\vspace{0.5cm}

\narrowtext

\end{@twocolumnfalse}
]

\section{Introduction}

Geometric frustration \cite{book} spans a wide range of 
systems- Frank-Kasper phases \cite{Frank-Kasper}
\cite{nelson},\cite{sadoc}, disordered materials, clathrates, and complex 
fluids such as the cholsteric blue phases \cite{sethna},
and amphiphilic membranes \cite{SC}. Perhaps most notable
of these materials are metallic glasses, and 
possibly glasses in general, whose ideal icosahedral
(and more complicated extensions)
packing is not compatible with Euclidean
space filling. Such packings have seen 
a revival in the study of quasicrystals. 
Till now many theoretical treatments, e.g. \cite{subir},
have treated geometrical frustration as a 
perturbative effect akin 
to the smooth suppression of the 
superconductivity by the 
insertion of a frustrating 
magnetic field and the
appearance of topological
defects (vortices) in type II
superconductors. A first 
hint that the physics
might be far richer 
and reward us with 
interesting surprises
was found in \cite{chayes}, \cite{joe}
where the insertion of 
an infinite range Coulomb
interaction was shown to
be non-perturbative.
By analogy to the 1/r gravitational 
potentials generated by
geometrical deformations,
all of this suggested 
that a similar occurrence
might arise in geometrically 
frustrated systems.

Here we will show how an explicit infinitesimal geometric frustration 
may lead to a dramatic, 
non-perturbative, effect on the thermodynamics in a multitude
of systems.
As a result of geometric frustration, the critical 
temperature might be discontinuously depressed and an ``avoided
critical point'' will emerge. The physical origin 
of the discontinuity is a greatly enhanced 
degeneracy: the geometric frustration thwarts 
ideal order (or crystallization) and leads to many competing 
low energy states. In this article we will focus on how 
this arises in various $O(n)$ systems immersed in uniform
$SO(n)$ backgrounds. 
Similar models \cite{sethna},\cite{subir} were 
previously employed to describe the microscopics 
of metallic glasses and correctly predicted the 
locations of the structure factor peaks \cite{subir}. We 
will show that that such models exhibit 
``avoided critical behavior'': a 
genuine true non-analyticity 
only at very low temperatures
as compared to the (avoided) critical 
temperature of the system sans 
any frustrating external 
non-abelian gauge background. 
Some of the best fits \cite{kivelson}, to date, 
to the dynamics of supercooled liquids 
were inspired by such notions for 
which we now provide a firmer 
microscopic basis. We will compare 
one of the fitted parameters
in these plots (the magnitude of 
the ``avoided critical temperature''
for each of the supercooled liquids) and demonstrate
that, on average, these might be well
predicted by our microscopic
picture.

To couch geometrical frustration in a general field
theoretical setting, we will consider systems 
whose continuum Lagrangian density may be
written in the form 
\begin{eqnarray}
{\cal{L}}_{\mbox{matter}} = \frac{1}{2} |D_{\mu} \phi^{\nu}|^{2} -
\frac{M^{2}}{2} |\vec{\phi}|^{2} + \frac{u}{4!} 
|\vec{\phi}|^{4} + ...,
\label{general}
\end{eqnarray}
with a covariant derivative $D_{\mu}(x) = \partial_{\mu} - 
i \theta A_{\mu}(x)$ 
appropriate for parallel transport on a surface 
whose curvature is $(1/R) = \theta$. As we will
detail below, the strength of the coupling to the 
background gauge field ($\theta$) sets the 
degree of geometric frustration present 
in the system. 

Let us quickly sketch what 
unfolds as the frustration is introduced. Matters 
become most transparent in momentum space
which diagonalizes the quadratic part 
of the action. When $\theta=0$, the 
quadratic part of the action has a
single minimum at wavenumber $\vec{k}=0$.
In this standard case, long wavelength
(small $|\vec{k}|$) fluctuations can 
destroy order in dimensions $d \le 2$
and indeed they do so for continuous 
fields $\vec{\phi}$. In high dimensions ($d>2$), 
order reasserts itself at temperatures 
$T \le T_{c} = {\cal{O}}(1)$.

\begin{figure}
\includegraphics[width=4.5cm]{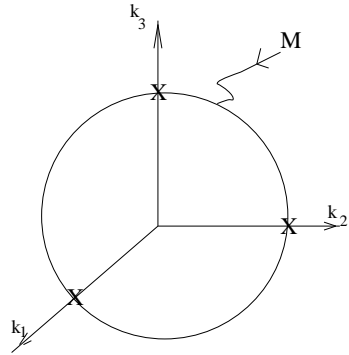}
\caption{A schematic of the interaction
kernel, $v(\vec{k})$, minima
in Fourier space. In all of the
systems that we consider
here, the relevant continuum limit 
$v(\vec{k})$ attains its minima on 
shells $|\vec{k}| = q >0$.
In the presence of 
lattice point group
symmetry terms, the
rotational symmetry
is lifted and the minima
appear at a discrete
number of points (labeled
by ``X'' in the figure).}
\label{manifold}
\end{figure}

This picture undergoes dramatic ramifications
as the frustrating field is switched on.
As seen, in Fig.(\ref{manifold}), the minimum in 
$\vec{k}$ space is no longer point like but 
rather extends over a (d-1) dimensional manifold 
whose radius is set by the geometrical 
frustration $\theta$. For any frustration
$\theta$, no matter how small, the manifold
of minimizing modes $M$ is a high dimensional
object spanning many modes. 
All Fourier modes on this surface are degenerate leading to
an exceptionally high configurational degeneracy in real
space. In the aftermath, an effective one dimensional
behavior results (which can be made
precise in the $n \to \infty$ limit) 
with no ordering transition in sight. 
Only at the unfrustrated point $\theta=0$ does
the minimizing manifold $M$ shrink to
the origin, and our old intuition regarding the 
long wavelength fluctuations is regained.

For simplicity, we will mostly 
consider a translationally invariant $D_{\mu}$
with non-commuting uniform gauge connections $A_{\mu}$.
Our results can be extended to systems with 
arbitrary non-uniform $D_{\mu}(x)$ in which
case our analysis may be repeated
with the Fourier basis index replaced 
by an index labeling the 
more complicated diagonalizing eigenbasis
(the minimizing manifold $M$ will merely
appear as an identical object in this
basis). 

We will consider non-Abelian 
gauge {\em background} fields $\{A_{\mu}(x)\}$ having
no dynamics of their own and
merely enforcing geometrical
frustration. To make the gauge field a true 
dynamical variable, we will 
need to add terms involving
derivatives; the simplest standard gauge invariant 
choice is the squared field
$(F_{\mu \nu})^{2}$ where the 
field $F_{\mu \nu} = [D_{\mu},D_{\nu}]$.  
The resulting well known Lagrangian 
\begin{eqnarray}
{\cal{L}}_{\mbox{matter + gauge}}=  
-\frac{1}{4} F_{\mu \nu} F^{\mu \nu} \nonumber
\\ + \frac{1}{2} |D_{\mu} \phi^{\nu}| ^{2}
- \frac{M^{2}}{2} |\vec{\phi}|^{2} + \frac{u}{4!}|\vec{\phi}|^{4}+...
\end{eqnarray}
Theories involving only the first term
are frequently referred to as ``pure gauge theories''. 
We will prove that in the opposite 
extreme of Eqn.(\ref{general}) (non-dynamic or small $A_{\mu}$)
a new surprising ``avoided critical behavior'' may occur:
the effect of coupling to the non-Abelian gauges
is non-perturbative. The theory suddenly becomes extremely 
sensitive to thermal effects. The density of
states becomes, to a certain extent,
one-dimensional. This also applies to some 
more general cases in which the effective action
after integrating out the gauge fields will
be of a form similar to Eqn.(\ref{general}).
For instance, if the kinetic term in 
the gauge fields is replaced by
$(F_{\mu \nu}- f_{\mu \nu})^{2}$ 
with specified frustrations $f_{\mu \nu}$ 
then after integrating out the gauge 
fields $A_{\mu}(x)$ the resulting 
action density in the matter
fields $\vec{\phi}$ will exhibit 
frustration. Such a theoretical 
prediction has serious implications
for many geometrically frustrated
systems which will form a new universality
class of their own. As we will aim
to highlight in the paper,
the very unusual ``glass
transition'' might be linked at its
very core to the non-perturbative 
physics spawned by  
non-Abelian background 
fields (although of a more
complicated form than 
of a fixed uniform
background).

\begin{figure}
\includegraphics[width=4.9cm]{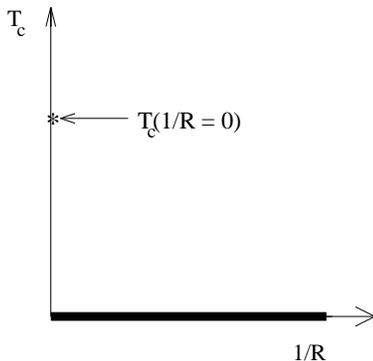}
\caption{The typical 
phase diagram encountered in $n \ge 2$ realizations
of our models. When the system is unfrustrated,
the critical temperature in dimensions $d>2$ 
is finite (and of order ${\cal{O}}(1)$).
When the minimizing manifold $M$ is a $(d-1)$ dimensional
shell, the critical temperature drops to zero.
In the above, 1/R corresponds to the inverse radius of the 
sphere on which ideal order (icosahedral or other) 
can prevail. 1/R scales
as $q$- the radius of the minimizing shell
$M$ is $\vec{k}$ space.}
\label{avoid}
\end{figure}

Our main conclusion in the continuum limit is schematically summarized 
in the large $n$ phase diagram of 
Fig.(\ref{avoid}). Here, the horizontal axis
denotes the strength $\theta \equiv R^{-1}$
of a uniform non-abelian $SO(n)$ gauge background.
The vertical axis denotes the critical 
temperature of a three dimensional 
$O(n)$ spin system. In the absence of 
a uniform non-Abelian background field
(i.e. at $\theta=0$), the spin system 
undergoes a phase transition at a finite 
$T=T_{c}(0)$ which is order of the 
exchange constant $J$. When a
non-abelian background field 
is applied, the system is unable to
order at finite temperatures.

We also report on a 
generalized Mermin-Wagner inequality
which allows us to connect the magnitude of
the order parameter (the absence
of entropic effects) to the 
characteristic relaxation
times present in any system
(including glass models).
We will also show how to derive 
Vogel-Fulcher dynamics
for a multitude of glass models
(including non-Abelian gauge
backgrounds), and try to fortify 
earlier ideas suggesting
avoided critical dynamics.

\section{Definitions and Conventions}
\label{notations}

Throughout, we will examine 
the interaction kernel $v$ appearing in 
the quadratic part of the 
Hamiltonian
\begin{eqnarray}
H = \frac{1}{2} \int \frac{d^{d}k}{(2 \pi)^{3}} ~ v_{ij}(\vec{k})
\phi_{i}(\vec{k}) \phi_{j}(-\vec{k}).
\label{H-g}
\end{eqnarray}
This is the Fourier transform of
the real space 
$H = \frac{1}{2} \int d^{d}x \int d^{d}y V_{ij}(\vec{x}-\vec{y}) 
\phi_{i}(\vec{x}) \phi_{j}(\vec{y})$.
For spins lattice systems we will 
consider Hamiltonians of the type
\begin{eqnarray}
H = \frac{1}{2} \sum_{\vec{x},\vec{y}}
V(\vec{x}- \vec{y})[\vec{S}(\vec{x}) 
\cdot \vec{S}(\vec{y})]. 
\end{eqnarray} 
The sites $\vec{x}$ and $\vec{y}$
lie on a (generally hypercubic)  
lattice of size $N$, and 
$\{S_{i}(\vec{x})\}$ denote the $i$-th component
of the spin $\vec{S}$ situated at a certain 
site $\vec{x}$. The normalized spins 
have $n$ components. Throughout, we employ the non-symmetrical
Fourier basis convention
($f(\vec{k}) = \sum _{\vec{x}} 
F(\vec{x}) e^{-i \vec{k} \cdot \vec{x}};$
$ ~ F(\vec{x}) = \frac{1}{N} \sum_{\vec{k}} 
 f(\vec{k}) e^{i \vec{k} \cdot \vec{x}}$) .
The vector $\vec{q}$ denotes a wave-vector minimizing the kernel
$v(\vec{k})$: $v(\vec{q}) = \min_{k} \{v(\vec{k})\}$.
In the usual ferromagnetic models $\vec{q} = 0$ is the 
only minimizing mode.
For simplicity, we set the lattice constant to unity
and consider a lattice (of size $L$) with
periodic boundary conditions. The wave-vector
components $k_{l} = \frac{2 \pi r_{l}}{L}$
where $r_{l}$ is an integer (and the real space 
coordinates  $x_{l}$ are integers).
In all systems that we study in this paper
the Fourier modes $\vec{q}_{i}$
minimizing the interaction kernel $v(\vec{k})$ 
lie on a (d-1) dimensional shell.
We often refer to this high dimensional 
minimizing manifold as $M$. The high dimensional $\vec{k}$ space 
redundancy leads to a complicated ``energy landscape'' for real 
space configurations. As there are many zero modes tangent to $M$ 
that cost no energy, the system is very fragile
to perturbations and in the large $n$ limit, $T_{c} =0$.
On a lattice, the 
continuous rotational symmetry
is lifted. Here,
we find a finite number of 
minimizing modes (denoted
by an ``X'' in Fig.(\ref{manifold})). 
Notwithstanding,  the critical temperature will typically be much
lower than that of the unfrustrated system with a
single minimum at $\vec{k}=0$. The critical temperature
will drop discontinuously (from $T_{c} = O(1)$ to 
$T_{c} \ll O(1)$) as $M$ expands from 
the origin.

\section{Icosahedral Order In Brief}

Unlike other sections to follow,
all presented here is a review
of earlier work \cite{book}, \cite{sethna}, \cite{subir} 
which inspired the current
publication. When a liquid is 
supercooled (when cooled so rapidly
that it may not 
crystallize), and cooled further down
until its relaxation times become effectively infinite
it becomes a ``glass''. In slow
solidification into an
ordered crystal, the atoms/molecules readjust their positions,
slowly ``computing'' their ideal crystalline 
positions and veer toward them. 
Now, consider
the situation, for the 
supercooled liquid, when the 
``computation'' needs to be 
done almost on the spot. 
Here, the atoms/molecules
can only probe their immediate surrounding
to minimize 
energy costs of bonds with 
their nearest neighbors. For a Lennard-Jones 
interaction, each atom would like to
be some distance $r_{\min}$ 
(the distance at which the Lennard
Jones like potential $V(r)$ is minimized)
away from all of its neighbors
in order to minimize each 
bond individually. Here,
crystallization is trivial
in two dimensions- the local 
minimum can be extended to tile
the entire plane. Locally and globally, 
the ideal packing is
hexagonal with each atom surrounded by its 
six nearest neighbors all equidistant
from it at the ideal distance $r_{\min}$.
The local minimum 
that the atoms may find is also the global minimum.
Such a fortunate occurrence does not
arise in three dimensions. Four atoms
may be equidistant from each other
to form the vertices of a perfect 
tetrahedron. However, around a given 
edge of the tetrahedron we may not  
pack five other tetrahedrons 
perfectly: We can almost
do so leaving a $7^{o}$ void.
Similarly, particles whose locally preferred 
structure is an icosahedron (of 13 particles)
cannot globally tile all of space as seen by
the ``illegal'' icosahedral five-fold rotational 
symmetry \cite{nelson-ico}.
If this is correct, 
then microscopically, local clusters should
appear. 
Certain supercooled metallic
glasses offer a 
good testing ground for these notions.
There is evidence that
short-range order in under-cooled 
liquids and metallic glasses is 
icosahedral \cite{phys_today}. A theoretical
construct invented long ago \cite{book,nelson,sethna} 
is a high dimensional 
reference system in which 
perfect local icosahedral order,
with equal bond distances,
could easily tile large regions 
without any voids. Endowing three dimensional 
space with a small curvature, the
$7^{o}$ void 
can be compressed to zero: Icosahedral clustering
is perfect in a slightly curved three
dimensional space.  An ideal, icosahedral crystal
on the surface of a 4-sphere 
(polytope ``$\{3,3,5\}$''),
can be found. It
consists of 120 particles
embedded on the slightly curved spherical 
surface $S^{3}$ (the three 
dimensional boundary of a sphere 
in four dimensions). 
We (\cite{kivelson})
have taken this
notion to the extreme and 
argued that the dynamics
of glasses is controlled
by the vicinity to the 
melting transition of 
ideal crystal that may 
form on the curved 
sphere. 
Regions of short-range $\{3,3,5\}$ order
in are broken up by an
array of $-72^{o}$ disclination lines, 
forced in by ``frustration''- the incompatibility
of flat space with a space filling 
icosahedral crystal. The Frank-Kasper phases 
of transition metal alloys are ordered arrays
of disclination lines in such an icosahedral medium.
The order parameters $Q_{n,m_{a},m_{b}}$ 
are obtained by projecting the local particle configuration
onto the hyperspherical 
harmonics $Y_{n,m_{a},m_{b}}$ (the spherical harmonics 
(``$Y_{l}^{m}$''-s) for a 4-sphere)
of a tangent four-sphere which can accommodate
icosahedral order \cite{subir}.

\begin{figure}
\includegraphics[width=5.0cm]{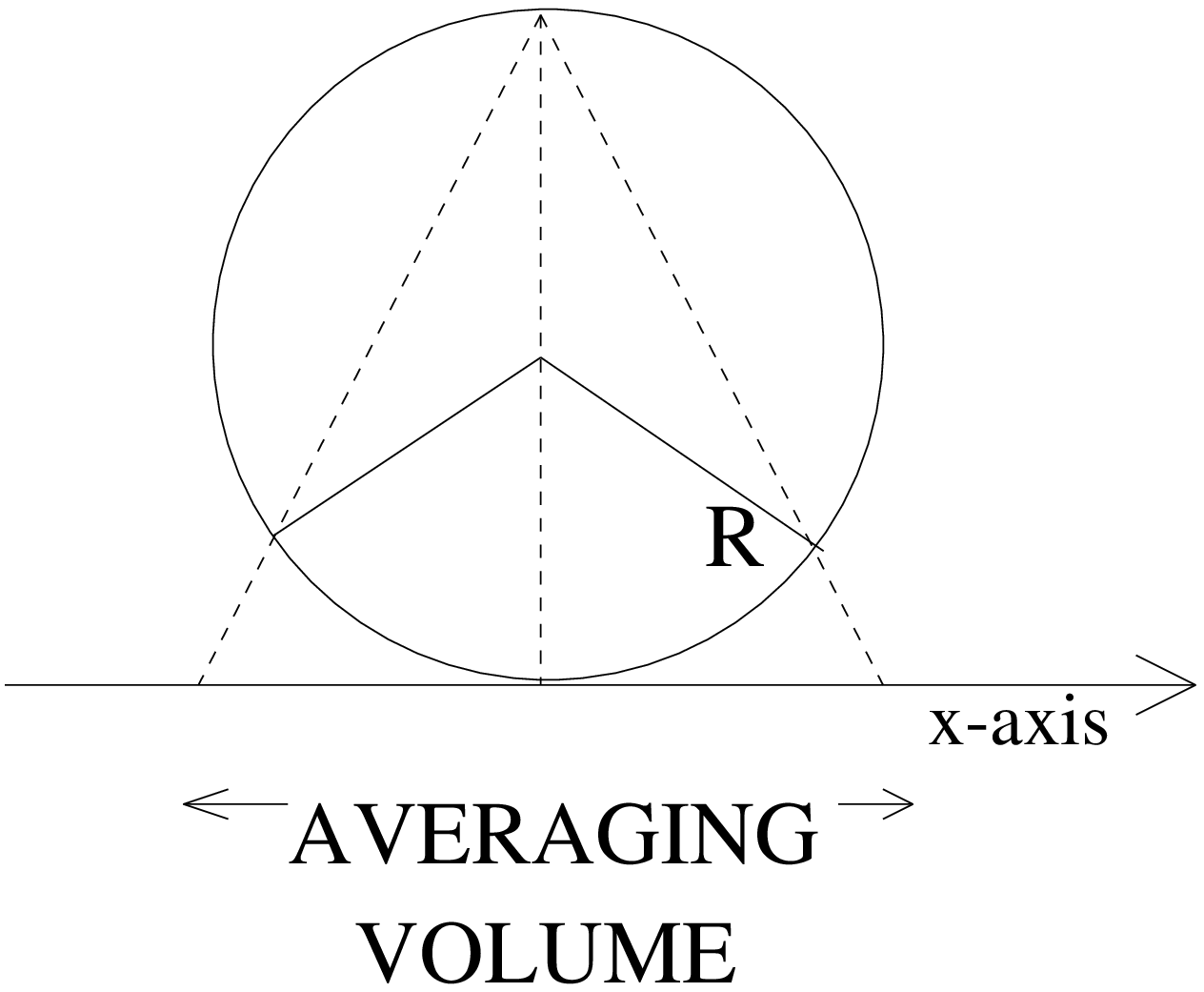}
\caption{A one dimensional analogue. We project the mass
density from the real-space $x-$axis 
onto a disk of radius $R$ (with internal
coordinate $\theta$) associated with each point on 
the $x$-axis and expand in harmonics
($\rho(x,\theta) = \sum_{m} \rho_{m}(x) e^{i m \theta}$).
Ideally, the mass density on the line matches that
obtained by rolling the disk and the local energy functional
$[\partial_{x} - \partial_{\theta}/R] \rho(x,\theta)]^
{2}$ is minimized. Integrating over $\theta$ we find
the one dimensional analogue of the $SO(4)$ action below.}
\label{}
\end{figure}

In three
dimensions, a 3-sphere is attached to 
every point $\vec{x}$ in 
real space. 
The mass density is projected onto
these spheres. The density at all
$\vec{x}$ on the tangent sphere ($\hat{u}$
is the internal coordinate on the sphere),
\begin{eqnarray}
  \rho(\vec{x},\hat{u}) \equiv \sum_{n;m_{+},m_{-}}
  Q^{*}_{n;m_{+},m_{-}}(\vec{x}) Y_{n;m_{+},m_{-}}(\hat{u}),
\end{eqnarray}
upon expansion in the hyper-spherical harmonics $Y_{n;m_{+},m_{-}}$. 
The $m_{\pm}$ subscripts in $Y_{n;m_{+},m_{-}}$ refer
to the two $SU(2)$ representations (of 
generators $\vec{A}_{\pm}$ of spin n/2)
that the $SO(4)$ representation can be decomposed into.
The Hamiltonian
\begin{eqnarray}
  H= \frac{1}{2} \sum_{n} [K_{n} |(\partial_{\sigma} - \imath \theta L_{o
    \sigma}^{n})\vec{Q}_{n}|^{2}+\mu_{n} |\vec{Q}|^{2}],
\label{SO-4}
\end{eqnarray}  
where we organize the $\{Q_{n;m_{+}m_{-}}\}$ multipoles for each
$n$ into $(n+1)^{2}$ dimensional vectors in the
$|n/2,m_{+};n/2,m_{-} \rangle $ basis. In a large $n$ analysis,
the second (mass) term portrays
the effect of
normalization.  
%The kernel is
%\begin{eqnarray} 
%  Q^{*}_{n}(\vec{k})[k^{2}- 2 \theta \vec{k} \cdot (\vec{M}_{+}
%  +\vec{M}_{-})+\nonumber
%\\ \theta^{2}(\vec{M}_{+}+\vec{M}_{-})^{2}]
%  Q_{n}(\vec{k})],
%\end{eqnarray}
%with $Q^{*}$ the Hermitian conjugate. 
Fourier transforming and diagonalizing \cite{subir},
\begin{eqnarray}
H = \frac{1}{2} \sum_{\vec{k},M,i,n}
|\alpha^{n}_{M,i}(\vec{k})|^{2}[K_{n}
\lambda_{M,i}^{n}(\vec{k})+\mu_{n}],
\label{free-SO-4}
\end{eqnarray}
with $M$ the eigenvalues of $\hat{k} \cdot [\vec{A}_{+} + \vec{A}_{-}]$,
the amplitudes of the i-th eigenmode $\alpha_{l}^{i}(\vec{k})$, 
and $\{\lambda_{M,i}^{n}(\vec{k})\}$
the eigenmode spectrum. 
Diagonalization yields, in all cases, a set
of eigenvalues \{$\lambda_{n}^{i}(|\vec{k}|)$\} with the global
minimum (for each $n>0$) occurring at $k = q_{n}^{i}>0$, in agreement
with structure factor measurements (displaying a series of peaks at
these minimizing wave-vectors). By rotational symmetry,
the eigenvalues depend on the magnitude $|\vec{k}|$
of the wavenumber. Thus, as the minima do not
occur at the origin, shells of minimizing modes
(as in Fig.(\ref{manifold})) occur.
We now define
\begin{eqnarray}
\min_{n,i,\vec{k}} \{K_{n} \lambda_{n}^{i}(|\vec{k}|)\} =
K_{n_{\min}}\lambda_{n_{\min}}^{i_{\min}}(q),
\end{eqnarray}
where $n_{\min}$ and $i_{\min}$ denote the 
values of these indices at which the global
minimum is attained- the corresponding
minimizing wavenumber is denoted by $\vec{q}$.
By rotational symmetry, all that matters
is the modulus $q$. As expected, 
$q$ scales with the geometric frustration 
$\theta$. 
The values of $n$ allowed by icosahedral symmetry
are $n=0,12,20,24,30,32,36$. The smallest non-trivial 
$n$ corresponds to a $(n+1)^{2}=169$ complex component vector
$Q_{n,m_{+},m_{-}}$. For a discussion on how these 
ideas may also be applied to wave functions 
on a sphere \cite{the_wave}.

\section{Non-perturbative Thermodynamics}

We will argue, via a large $n$ analysis, a Mermin-Wagner inequality
for similar models, and a fluctuation
analysis \cite{thesis},\cite{me}, that these and 
other {\em continuum} 

\[ \fbox {glass\, models\, do\, not\, order\, at\, finite\, temperatures.}\]

Any finite frustration $\theta$, no matter
how small, drives these systems
away from criticality at finite temperatures.
In many solvable cases, this 
non-perturbative effect has nothing
to do with the breakdown of perturbation
theory due to long range interactions.
For any value
of the frustration
$\theta$, the minimizing manifold
of modes in Fourier space 
is high dimensional.
The high dimensionality allows
for the proliferation of many zero energy modes which 
easily erase any sign of potential
order. When small point 
group symmetry terms lift the
rotational symmetry of continuum
models, the critical temperature becomes 
finite yet still minute as compared 
to the ordering temperature of
the unfrustrated system,
\begin{eqnarray}
{\cal{O}}(1) = T_{c}(\theta=0) \gg T_{c} (\theta  \neq 0). 
\end{eqnarray}

\section{The Large $n$ Phase Diagram}

Whenever the minimizing mode $q$ is finite and a
(d-1) shell of minimizing modes occurs (as in 
Fig.(\ref{manifold})), a large $n$ calculation reveals 
that the {\em critical temperature},
$T_{c}^{Spherical}=0$. As a function of
the radius of the minimizing shell $M$
in Fig.(\ref{manifold}), this drop
occurs {\em discontinuously}.

In the large $n$ (or spherical) limit, we subject the 
system to a normalization constraint of
the $(n+1)^{2}$ (that of $n=n_{\min}$)
component vector $\vec{Q}$,
\begin{eqnarray}
\sum_{M,i,\vec{k}} |\alpha_{M,i}^{n_{\min}}(\vec{k})|^{2} =1.
\end{eqnarray} 
We solve the system of 
Eqn.(\ref{free-SO-4}), subject to
the latter (spherical or large $n$)
constraint, to find
that the critical temperature,
\begin{eqnarray}
\frac{1}{k_{B}T_{c}} = \sum_{M,i} \int \frac{d^{3}k}{(2 \pi)^{3}}
\frac{1}{K_{n_{\min}} \lambda_{M,i}^{n_{\min}}(\vec{k})- K_{n_{\min}}
\lambda_{0,i_{\min}}^{n_{\min}}(q)}. 
\label{dramatic}
\end{eqnarray}
As all terms in the integrand are explicitly positive,
we may bound the right-hand side by its contribution
from $(i_{\min},M=0)$. If the $(i_{\min},M=0)$ eigenvalue is 
analytic near its minimum at $|\vec{k}|=q$, then using the
shorthand  $F(|\vec{k}|) \equiv K_{n_{\min}}
\lambda_{0,i_{\min}}^{n_{\min}}(\vec{k})$,
a trivial bound (from the region $[q-\delta,q+\delta]$)
reads
\begin{eqnarray}
\frac{1}{k_{B}T_{c}} > 
\frac{(q- \delta)^{2}}{\pi^{2}}
\int_{q-\delta}^{q+\delta} dk 
\frac{1}
{[F^{\prime \prime}(q)] (k-q)^{2}}
\label{one-dim+} 
\end{eqnarray}
The integral in 
Eqn.(\ref{one-dim+}) diverges.
Here we assumed that the second 
derivative of $F$ is finite at $k=q$;
if the second derivative vanishes then 
the divergence of the right-hand side of 
Eqn.(\ref{dramatic}) is even more dramatic.
{\em The divergence signals that the critical
temperature $T_{c}=0$}.
Critics might
argue that this might be
an artifact of the large
$n$ approximation
and that, ``for well known''
reasons, this or that may 
happen (e.g. fluctuation driven 
first order phase transitions \cite{Brazovskii}). 
To counter such arguments, we prove 
in a later section
a generalized Mermin-Wagner  
theorem of related rotationally
symmetric spin models disallowing magnon
dispersion relations about any 
ordered system
at arbitrarily low temperatures $T=0^{+}$!
Even if a fluctuation driven first order transition 
occurs, as we show (by extending replica calculations),
the system may be frozen in a glass state before reaching
the transition, so its existence might be immaterial.
For the moment, we simply note that the lowest lying 
vector $\vec{Q}_{n=12}$ has 169 complex components.
This number is large and a $1/n$ expansion
seems plausible. In contrast to certain expectations, it is seen that 
many beautiful early constructs \cite{book,subir,sethna}
predict a vanishing $T_{c}$ for any finite
non-Abelian coupling if no symmetry breaking terms are present. 
As shown in Fig.(\ref{avoid}), the
phase diagram corresponding to these ideal metallic glasses, and
plausibly to dense random packed supercooled liquids in general, shows
an ``avoided critical point''- a spike for tunable $\vec{q}=0$ in the
interaction kernel, a genuine non-analyticity in the (thermo)dynamic
functions at finite coupling to the uniform
gauge background only at $T_{c}=0$.

\section{Avoided Criticality via 1/n}

\begin{figure}
\includegraphics[width=5.5cm]{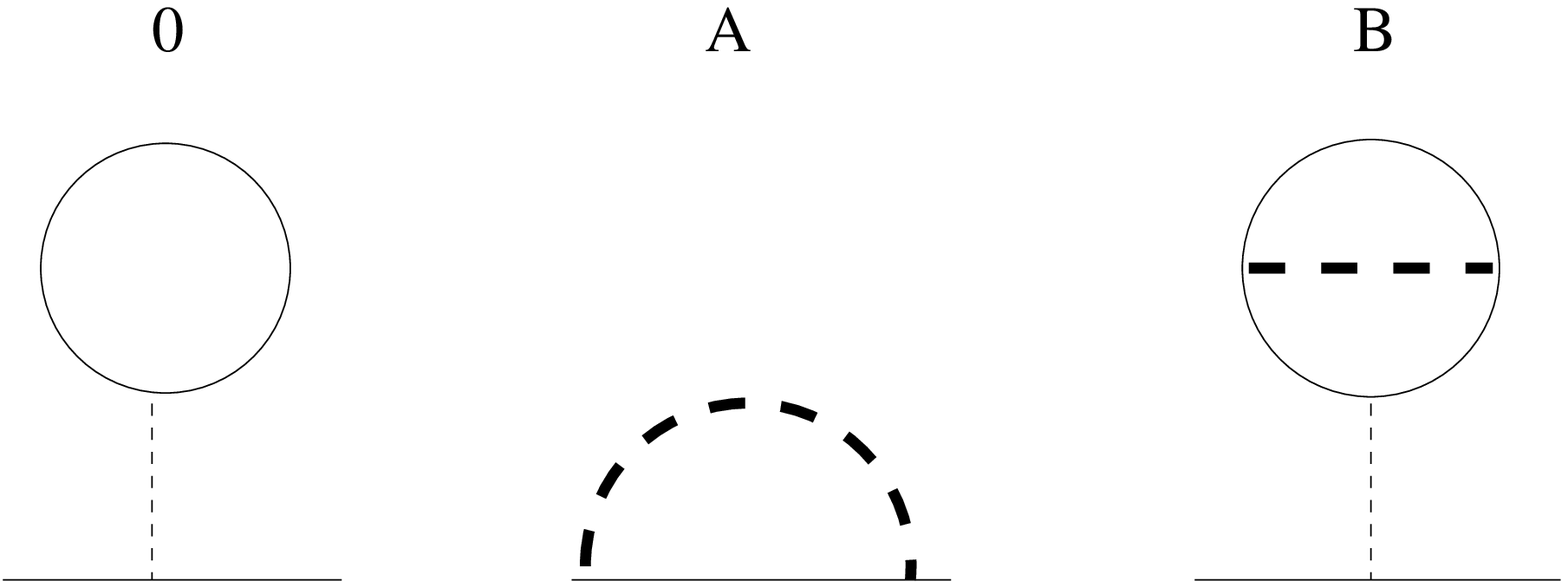}
\caption{The self energy corrections. The
thin dashed lines denote bare interactions.
Thick dashed lines represent dressed interactions
(i.e. the bare interactions screened
by a geometric series of bubble
diagrams). The solid
lines denote propagators.
$\Sigma^{0}$ is the $O(1)$ parquet
diagram. $\Sigma^{(A),(B)}$ are $O(1/n)$
corrections.}
\label{largen}
\end{figure}

Within a $1/n$ diagrammatic framework, the absence of 
a finite temperature phase transition, in the large
$n$ limit, is seen by the divergence of 
the lowest order (tadpole) contribution
to the self energy (Fig.\ref{largen}).
With the self-consistent
propagators,
\begin{eqnarray}
G^{-1}_{i_{\min},n_{\min}}(\vec{k}) = 
K_{n_{\min}} 
\lambda_{0,i_{\min}}^{n_{\min}}(\vec{k}) +
r 
\end{eqnarray}
for the lowest energy mode ($n_{min},i_{min}$) of
relevance at low temperatures, with the 
bare propagator
\begin{eqnarray}
G^{-1}_{0}(\vec{k}) = 
K_{n_{\min}} 
\lambda_{0,i_{\min}}^{n_{\min}}(\vec{k}) +
r_{0}. 
\end{eqnarray}
The zeroth order self-energy
is a momentum independent constant
\begin{eqnarray}
\Sigma^{(0)} = \int \frac{d^{3}k}{(2 \pi)^{3}}
G_{i_{\min},n_{\min}}(\vec{k}).
\label{self-z}
\end{eqnarray}
The zeroth order self-energy contributions from all other 
(non-$(n_{min},i_{min}$)) eigenvalues are strictly 
positive. When the
mass gap vanishes, i.e. $r_{\min} = - K_{n_{\min}} 
\lambda_{0,i_{\min}}^{n_{\min}}(q)$,
the integral in Eqn.(\ref{self-z}) diverges. 
By the Dyson equation,
\begin{eqnarray}
G^{-1}(\vec{k}) = G_{0}^{-1}(\vec{k}) + \Sigma(\vec{k}).
\end{eqnarray}
A divergent $\Sigma^{(0)} \sim (r-r_{\min})^{-1/2}$ signals the inability of 
the mass gap to vanish. Only at $r_{0} = -\infty$
can $r=  - K_{n_{\min}} \lambda_{0,i_{\min}}^{n_{\min}}(q)$.
In the conventional (finite $T_{c}$) case,
the bare mass $r_{0} = a(T-T_{0})$ codes for a linear temperature scale.
The divergence that we obtain here implies that
$T_{c}=0$ and that consequently a linear temperature 
scale is void in its environs. A similar
analysis for the quantum case is detailed
in the appendix. There are two first order 
1/n diagrams and, after regrouping
diagrams self-consistently, six important second order diagrams). 
By re-summing \cite{thesis} diagrams self-consistently, the zeroth 
order divergence forbidding finite temperature ordering 
may stay in tact.

\section{New $SO(n)$ background spin models}

Here we, by inventing a simple ``soccer-ball model'', we
intuitively illustrate a simple basic premise:

\bigskip

{\em A non-Abelian background leads to High
Dimensional minimizing shells in $\vec{k}$-space.}

\bigskip

The radius of the shell is set by the strength
of the non-Abelian background. The profound increase in the dimensionality
of the minimizing modes leads to a highly
non-perturbative effect on the thermodynamics.
In momentum space, there are many zero-energy, ``Goldstone'' like, 
modes tangent to the minimizing manifold. These 
fluctuations lead to a vanishing spin stiffness.
All of this is in stark contrast to the innocuous effect of abelian
gauge backgrounds: A $U(1)$ minimal substitution 
$v(\vec{k}) = (\vec{k}-e\vec{A})^{2}$, 
with a constant $\vec{A}$, trivially shifts the point minimum in
$\vec{k}$ - space. For
non-Abelian coupling the effect is much more subtle:
{\em the minimizing manifold is a shell spanning 
an infinite number of Fourier points.} 
As the original models of \cite{subir} only allow
for representations of SO(3) or SO(4) 
with a high number of components (large n indeed)
for which calculational results may not be too transparent, 
we will now invent a simple toy spin model to gain intuition:
the ``soccer-ball model''. The interplay between the external 
spatial and internal spin direction that it
possesses is present in very
few other spin systems
of a very different
character (e.g. orbital Jahn-Teller
and Kugel-Khomskii like compass models). 
Consider three component
(Heisenberg) spins on the square lattice with SO(3) couplings:
\begin{equation}
  H = - \sum_{ \langle x,y \rangle} \vec{S}(\vec{x}) * \vec{S}(\vec{y}).
\end{equation}
Here the $*$ operation denotes to the scalar product between a spin
$\vec{S}(\vec{x})$ attached to the 
south pole a sphere (a ``soccer-ball'') and
subsequently rolled to the neighboring site $\vec{y}$ and the
spin $\vec{S}(\vec{y})$ situated at $\vec{y}$ ; the action is locally
minimized by a spin rotation in the $[\ell n] \mbox{ } plane $ for a
lattice displacement along the $\ell-axis$.  For a rolling angle
$\theta$, we will now consider a spin configuration having as its 
only non-vanishing Fourier
components $\vec{S}(\pm \theta \hat{e_{1}})$, 
namely, the configuration
\begin{eqnarray}
(S_{1}, S_{2}) = (\cos \theta x_{1}, \sin \theta x_{1}) 
\end{eqnarray}
with $S_{1}$ and $S_{2}$ 
the x and y components of the spin and $x_{1}$ the spatial x
coordinate. This spiral configuration
will saturate all bonds along $\hat{e}_{1}$, and satisfy
$\vec{S}(\vec{x})=\vec{S}(\vec{x}+\hat{e}_{2})$ (just as for the uniform
($\vec{k}=0$) configuration). The energy of the above configuration is
lower than that of a $\vec{k}=0$ mode. In the continuum limit,
the kernel is invariant under rotations of
$\vec{k}$. As the minimum does not occur at the origin, a shell of
minimizing modes is seen to exist (a $(d-w)$ dimensional manifold of
minimizing modes is generated for $O(n=d+2-w)$ spins with the above
``rolling'' $SO(n)$ action for all but $(w-1)$ directions assigned
ferromagnetic couplings). Alternatively, in the continuum limit, the
differential action evaluated for the above configuration along any
ray not parallel to $\hat{e}_{2}$ is lower than that for the uniform
configuration.  The continuum Hamiltonian
\begin{equation}
  H = \frac{1}{2} \int |D_{\mu} S^{\nu}|^{2} d^{2}x = \frac{1}{2}
\int S_i(\vec{k})
  v_{i,j}(\vec{k}) S_{j}(-\vec{k}) d^{2}k,
\end{equation}
a nonlinear $\sigma$ model
on a curved surface.
The covariant derivative
\begin{equation}
  D_{\mu} = \partial_{\mu} +\imath \theta \epsilon_{\mu \nu}
  \hat{L}_{\nu},
\end{equation}
where $L_{\nu}$ is the $\nu$ component of the $\ell=1$ angular momentum
generator ($\epsilon_{12} = -\epsilon_{21} = 1 $). Demanding that
the ``rolled'' vector increment be real suggests the angular momentum
representation $(\hat{L}_{i})_{jk} = -\imath \epsilon_{ijk}$. 
The interaction kernel
pursues the form:
\begin{equation}
  v_{i,j}(\vec{k}) = k^{2} \delta_{i,j} + 2 \theta \epsilon_{\mu \nu}
  k_{\mu}(L_{\nu})_{i,j} + \theta^{2}(L_{x}^2+L_{y}^2)_{i,j}.
\end{equation} 
The eigenvalues are invariant under planar rotations of $\vec{k}$.
The minima occur on the ((2-1)dimensional) shell:
\begin{equation}
  \vec{q} \in M_{\theta}: \vec{q}\mbox{ }^{2} = A \theta^{2} .
\end{equation}

For the lattice version $v_{ij}(\vec{k})= \sum_{\pm \ell} e^{\pm
  \imath k_{\ell}}R_{\ell}(\pm \theta)$ where $R_{\ell}(\pm \theta)$
is the rolling matrix (rotation matrix about the $\ell` \neq
\ell$-axis ).
The continuum representation $L_{i} = -i \epsilon_{ijk}$ (for the
change of the spin $\delta \vec{S} = \delta \vec{\phi}$ x $\vec{S}$ 
incurred by a rotation of angle $\delta \vec{\phi}$) emerges
trivially.

Exploiting rotational invariance in the continuum limit, we may
always rotate a pure mode $\vec{q}$ of unit $(O(n))$ spins such that
it lies parallel to $\hat{e}_{1}$. All bonds parallel to $\hat{e}_{2}$
will entail the same energy penalty as the $\vec{k}=0$ mode and to
minimize energy cost along $\hat{e}_{1}$ we simply set $\vec{q} =
\theta \hat{e}_{1}$. We now relax the constraints of locally
normalized real spins  and allow the spins to be complex and 
only satisfy a global normalization constraint. 
Any ground states found here will obviously be of equal or
lower energy than those of three-component Heisenberg spins. 
In a spherical model of complex spins with global normalization: $
\sum_{\vec{x}}|\vec{S}(\vec{x})|^{2} = N$ enforced by a Lagrange
multiplier $\mu$, the continuum limit kernel reads $$
\pmatrix
{k^{2}+\theta^{2}+\mu & 0 & 2 \imath \theta k_{y} \cr
  0&k^{2}+\theta^{2}+\mu&2\imath\theta k_{x}\cr -2\imath\theta k_{y}&
  -2 \imath\theta k_{x}& k^{2}+2 \theta^{2}+\mu\cr}.$$
Due to an inherent coupling between real and imaginary components,
 $|\vec{q}| = \theta$ may not be enforced. 

As they must, all eigenvalues degenerate onto the
canonical $k^{2}$ dispersion in
the limit $\theta \to 0$.
Two of the eigenvalues 
\begin{eqnarray}
\lambda_{o}^{\prime} \equiv \lambda_{o} + \mu = 
k^{2}+\theta^{2}+\mu, \nonumber
\\ \lambda_{\pm}^{\prime} \equiv \lambda_{\pm} +\mu
= \mu+ k^{2}+\frac{3}{2} \theta^{2}\pm \frac{1}{2}
\sqrt{16 k^{2}\theta^{2}+\theta^{4}},
\end{eqnarray}
 are non-analytic in $|\vec{k}|$.
The global minimum  
\begin{eqnarray}
\min_{\vec{k}}\{\lambda_{-}^{\prime}\} = \frac{7}{16} \theta^{2}+\mu
\end{eqnarray}
occurs on the (2-1) dimensional annulus 
$|\vec{q}|=\frac{\sqrt{15}}{4} |\theta|$.
The corresponding eigenvector is 
\begin{eqnarray}
|u_{-} \rangle~ = ~{\cal{N}}~ |1, \frac{-2 i \theta k_{x}}{- \theta^{2}/2 + 
\frac{1}{2} \theta \sqrt{16 k^{2} + \theta^{2}}}, \nonumber
\\ \frac{-2 i \theta k_{y}}
{ - \theta^{2}/2 + \frac{1}{2} \theta \sqrt{16 k^{2}+
\theta^{2}}}\rangle  \nonumber
\\  \equiv 
|s_{1;-}^{R},s_{2;-}^{I},s_{3;-}^{I} \rangle
\label{u_}
\end{eqnarray}
where the $R,I$ superscripts denote real/imaginary components respectively.
The minimum of the non-oscillatory ``ferromagnetic'' 
mode occurs at the origin ,
\begin{eqnarray}
\min_{\vec{k}}
\{\lambda_{o}^{\prime}\}=\lambda_{o}^{\prime}(k=0)=\mu+
\theta^{2}.
\end{eqnarray}
A Landau-Ginzburg type analysis
(\cite{thesis}, \cite{me}) shows strong (non-logarithmic) divergent
fluctuations from $\vec{k} \sim \vec{q}~$  in directions (within
the 6-dimensional parameter space
($s_{1}^{R,I},s_{2}^{R,I},s_{3}^{R,I}$)~)  transverse to the 
eigenvector of Eqn.(\ref{u_})
(such that the norm $|\vec{s}|$ does not change to lowest
order). The generalization of our ``soccer-ball'' model
to SO(4) couplings on a three dimensional
lattice is discussed in Appendix(\ref{SO_4_SOCCER}).
The reader might be dismayed
that here we analyzed
a situation in which
the spins are complex. 
Here we attempt 
to capture as best
as possible the
complicated non-Abelian 
theories of glasses.
In the actual high
dimensional multipole 
expansion as in 
Eqs.(\ref{SO-4})
the ``spin components'' $\vec{Q}$
are complex.

%\section{Cholesteric Blue Phases}
%
%
%The blue phases are described by the
%Hamiltonian \cite{sethna}
%\begin{equation}
%H [\hat{n}] = \frac{1}{2} \int d^{d}x  ~K (D_{i}n)_{j} (D_{i}n)_{j},
%\end{equation}
%where 
%\begin{equation}
%(D_{i}n)_{j} = \partial_{i}n_{j} + w  \epsilon_{ijk} n_{k}
%\end{equation}
%with $\hat{n}$ the director (molecular axis 
%of the choleseteric molecules). The
%minimizing modes occur 
%on a shell of radius $q>0$.
%Our large $n$ analysis 
%(less justified here) 
%may be reproduced here to 
%yield the same thermodynamics.

\section{Other Interactions}

Non-Abelian background fields are not the only 
potential sources of glass formation and 
amorphous ``clump'' configurations. 
Many more physical Euclidean space
potentials seem to do the trick.
All of these interactions seem 
to share the generic feature 
that in Fourier space- the minimizing 
modes form a high dimensional shell (Fig.(\ref{manifold}))
leading to an extremely high degeneracy in 
real space. This in turn leads to
a ``complex energy landscape''
with many competing minima.

\subsection{The Coulomb Frustrated Ferromagnet}

My collaborators and I have
\cite{kivelson} modeled
the geometric frustration felt 
by molecules as they attempt to 
pack into a locally preferred structure 
by a long-range coulomb interaction 
forbidding the ideal local structure to
tile all of space. In the 
continuum limit, the Hamiltonian 
\begin{eqnarray}
H= \int d^{3}x [\frac{1}{2}(\nabla \phi)^{2}  + \frac{m_{0}^{2}}{2} \phi^{2} +
\frac{u}{4!} \phi^{4}] \nonumber
\\ + \int d^{3}x \int d^{3}x^{\prime} \phi(x)
V(x-x^{\prime}) \phi(x^{\prime}),
\label{Coul}
\end{eqnarray}
with the potential $V(r) = e^{2}/r$. 
The minimizing modes 
occur on shells as in
Fig.(\ref{manifold}) 
with $|\vec{q}| \simeq e^{1/2}$.
Here we will derive such a 
model from the underlying microscopics.

\subsection{Kac-type Potentials}

These two-body interactions,
\begin{eqnarray}
V(|\vec{x}|) = V_{0} \gamma^{d} \phi(\gamma |\vec{x}|)
\label{kac-p}
\end{eqnarray}
with the step function $\phi(y)=1$ for $y \le 1$ and
zero otherwise and $V_{0} = \alpha^{2}$ \cite{loh}. The constant 
$\alpha$ controls
the integral strength of the potential and 
$\gamma$ dictates the range of the potential.
We note that, physically, this potential corresponds to 
a weak core interaction inside a sphere
of radius $\gamma^{-1}$.
These interactions may model 
glass formation\cite{klein}. 
The Fourier transformed interaction is minimized at 
finite wave-numbers (as easily seen,
it has a maximum at $k=0$ when
all the Fourier phases add 
coherently). The Fourier transform of Eqn.(\ref{kac-p}),
\begin{eqnarray}
v(|\vec{k}|) = 4 \pi V_{0} \gamma^{3} [ \sin(k/\gamma)- (k/\gamma) \cos
(k/\gamma)]/k^{3}.
\label{F_T}
\end{eqnarray}
This kernel displays its minimum at $k=q>0$ with $q/\gamma \simeq
5.76$. The minimum indeed occurs on a $(d-1)$ dimensional shell such 
as that drawn in Fig.(\ref{manifold}). The kernel 
$v(k)$ is analytic near $|\vec{k}|=q$ allowing
us to Taylor expand about the minimum. 
By superposing Kac-type potentials, 
we may generate discrete, ``digitized'', versions
of the Lennard-Jones potential. For instance,
\begin{eqnarray}
V(\vec{x}) = \sum_{i} V_{i} \gamma_{i}^{d} \phi(\gamma_{i} |\vec{x}|),
\end{eqnarray}
with $V_{0}= \alpha^{2} \gg -V_{1} > 0$, $\gamma_{1} = 1/r_{1}$,
and $\gamma_{2} = 1/r_{2}$ corresponds to a
potential $V(\vec{x})$ which is $(V_{0}+V_{1}) \gg 0$ inside
a sphere of radius $r_{1}$ and is $V_{1} <0$ for $r_{1} < r \le r_{2}$,
and vanishes for all $r>r_{2}$. Superposing
the Fourier transformed kernels of Eqn.(\ref{F_T}),
shows a minimum on a spherical
shell of finite radius $|\vec{k}|=q>0$.  
Although the absolute minimum of a particles
interacting via a Lennard-Jones is an FCC crystal,
whenever the liquid is cooled rapidly it effectively
sees the radially symmetric finite radius 
Fourier modes as the minimizing modes.

\subsection{Z1 and Z2 potentials}

\begin{figure}
\includegraphics[width=6.0cm]{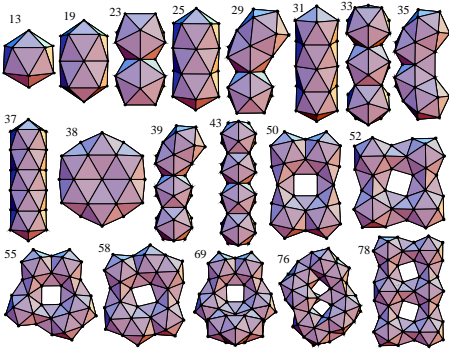}
\caption{The minimum energy clusters of the Z1 minima. The number
next to each cluster denotes the number of atoms in it.
This figure is reproduced from Doye et al.}
\label{JON}
\end{figure}

Lately, two new potentials for monatomic glass like
formers in the bulk were found \cite{jon}. In their presence,
the atoms display non-compact arrangements of 13-atom
icosahedra and a hint of fragile glassy dynamics in 
simulations of the cooled liquid
state. These are \cite{jon}
\begin{eqnarray}
V(|\vec{r}|)  = a \exp[- \alpha r]/r^{3} \cos(2 k_{F}r) + b(\sigma/r)^{n} +
V_{0},
\end{eqnarray}
with certain numerical values for the 
various parameters $\{a, \alpha, k_{F}, \sigma, n, V_{0}\}$
for $r>r_{c}$. This model can be related to
a superposition of Kac-type potentials. As in all
glass models that we investigate, the minimizing
Fourier modes of $v(\vec{k})$ occur 
on shells. Like other potentials, $V(r)$ naturally leads to
icosahedral like ordering over 
large non-compact regions.  
In the continuum limit, 
these potentials are akin
to non-Abelian
backgrounds.
The extended space 
that they occupy 
offers a genuine
example (not merely
that of small 13 particle
clusters) of relatively 
large systems ($\sim 10^{2}$) 
with geometric frustration.
The local maxima of the
potential inhibit closed
packed structures and
lead to polytetrahedral
structures which may be
divided into tetrahedra
with atoms at their
corners. Local icosahedral order involves
no disclinations but as the
clusters enlarge, negative disclinations
must be introduced. The disclination
lines cannot end abruptly and
must form closed loops 
or traverse the cluster. As in Frank-Kasper phases, 
the packing can be described by
a network of disclination lines
threading an icosahedrally coordinated
medium. Clusters of various sizes are depicted 
in Fig.(\ref{JON}) from \cite{jon}.  
For the two potentials for the largest 
clusters, networks and chains of icosahedra are 
respectively preferred. 
Upon cooling, {\em in silico}, a liquid of $\sim 10^{4}$ particles, 
extended icosahedral
clusters form with long time for local rearrangements. These potentials 
provide a natural realization
of the notions underlying
geometric frustration (here with icosahedral like
order over {\em a large range}), and a naturally related
glassy dynamics (slow rearrangement of 
icosahedral aggregates inhibiting
crystallization leading
to better glass-formers).

\section{Correlators And Disorder Lines}

\subsection{The $n \to \infty$ limit}

The correlation functions  (and much of the physics) are
governed by the ``thermal'' 
dynamics of the zeros and branch cuts of 
the energy spectrum in the 
complex k plane. Generically, in a system with many
competing interactions- with either
direct competition (as in the Coulomb
frustrated ferromagnet) or
indirect (e.g. non-Abelian
backgrounds)- the Green 
functions will display numerous
poles and/or branch points. 
These singularities,
generally, adiabatically shift their location
in the complex plane as a function
of temperature (the chemical
potential $\mu$ or mass 
gap $r$). Exactly how they 
evolve in the complex $k$ plane
determines, to a large extent, 
the nature of the transitions and 
crossovers that occur. 

In all of the models that we consider- 
all models with a minimum
of the kernel on a shell in $|k|$ 
space- we find that at low temperatures
the branch point or pole of relevance
has a finite real part leading 
to oscillations. This is expected as
at low temperatures, the relevant 
Fourier modes are on the shell
$|\vec{k}|=q$ and the system
must display oscillations.
As temperature is increased, the modulation
length becomes longer and longer.
The evolution of the modulation length
can be either continuous or discontinuous.
For systems with finite range interactions
(including non-Abelian backgrounds), the 
number of poles and branch-points is 
conserved as a function of temperature-
no two (or more) singularities merge
in the complex plane. In such systems,
the modulation length may, at times,
jump from one value to another.
This occurs when one pole
or branch point approaches the real 
axis more than another singularity
(and consequently has a smaller imaginary part).
The relevant asymptotic modulation length
then jumps to the singularity which 
is closer to the real axis. 
In our toy SO(3) actions, this indeed happens:
At high temperatures, the system is, asymptotically
unmodulated. Below a crossover temperature,
the system jumps and starts to exhibit finite
modulations associated with singularities of 
another eigenvalue. Naively, one might expect the branch
points present in the eigenvalue spectrum to immediately 
lead to algebraic (i.e. non-exponential) decay of 
correlations at all temperatures. This, however, is 
not what occurs. Cancelations arise 
and correlations decay exponentially 
at all finite temperatures (as expected).

In systems with competing interactions of different 
longer range interactions, the poles merge
frequently. For instance,  in the Coulomb Frustrated Ferromagnet,
the poles move along the imaginary axis at high temperatures.
At a certain crossover temperature, they merge and hit the 
``unit'' circle. From there, the poles bifurcate in pairs
to form two conjugate pairs of poles on the unit circle
leading to finite modulations which become of shorter and
shorter length (and ever larger correlation
length) as the temperature is decreased. The merger of
the poles at the crossover temperature leads to a
``disorder line'' like effect. The thermodynamic
functions are analytic at this temperature
yet their explicit real functional form changes
(just an analytic continuation from one natural 
set of functions to another). In systems 
with long range interactions (e.g.
the Coulomb Frustrated Ferromagnet of Eqn.(\ref{Coul}))),
the modulation lengths tend to 
decrease as temperature is decreased;
the converse occurs for the canonical interactions
($k^{2}$) thwarted or frustrated by an even shorter
range interactions (e.g. $k^{4}$).
In physical terms, the crossover (disorder line)
temperature is related to a temperature that marks the
gradual crossover from a one dimensional system
(the relevant density of states at low temperatures
is indeed one dimensional) to a canonical three 
dimensional system at high temperatures.
For explicit expressions for the 
correlation and modulation
length crossovers the reader us referred
to Appendix(\ref{explicit_crossover}).
A discussion of the thermodynamic
crossovers appears in Appendix(\ref{thermo_di}).

\subsection{A Josephson like length}

At very low temperatures the real space correlation 
function $G(\vec{x})$ decays differently in different regimes.
A $O(1/n)$ calculation reveals an algebraic cross-over similar 
to an inverted Josephson like regime. Specifically,
the pair correlator
\begin{eqnarray}
G(\vec{x}) \sim |\vec{x}|^{-2}~ ~ \mbox{for} \ell_{J}~ \ll |\vec{x}| \ll
L_{D} \ll \xi, \nonumber
\\ 
G(\vec{x}) \sim |\vec{x}|^{-1} ~ ~ \mbox{for} ~|\vec{x}| \ll \ell_{J},
\end{eqnarray}
with the Josephson like length
\begin{eqnarray}
\ell_{J} \equiv n/(2 \Sigma^{(0)}).
\end{eqnarray}
Evaluating matters explicitly (\cite{joe},\cite{thesis}),
we find that the Josephson like correlation
length $\ell_{J}$ scales 
as the inverse of the correlation 
length in the frustrated system.
Here $\Sigma^{(0)}$ is the ${\cal{O}}(n^{0})$ self-energy,
$\xi$ is the correlation length, and $L_{D}$, the domain
size, is $(2 \pi/|\mbox{Re}\{k_{-}\}|)$, with ``Re''
denoting the real part. More generally. within the 
general SO(4) theory, $k_{-}$ is replaced 
by the root of the eigenvalue corresponding 
the largest correlation length (smallest
imaginary part). This length scale
appears for all actions in which
the relevant interaction kernel
is peaked about a (d-1) dimensional
manifold $|\vec{k}|=q>0$ about which
they are quadratic.

\section{A Generalized Mermin-Wagner Inequality}

Here we report on an extension of the Mermin-Wagner
theorem which naturally links the magnitude
of the order parameter to constraints
on the characteristic decoherence 
or relaxation time. We generalized the Mermin-Wagner
inequality to all two dimensional systems
with an analytic interaction kernel
and to a host of {\em three dimensional
systems} in which a certain 
integral diverges. The inequality 
that we derive
reads 

\[ \fbox {$2|m_{q}|^{2}T \langle \tau_{\vec{k}} \rangle \le 1$}\]

with $m_{q}$ the magnetization at the minimizing momentum $q$,
and (with $\hbar =1 $) $\tau_{\vec{k}}$ is an integral having the 
dimensions of time reflecting the characteristic decoherence 
or relaxation time in a quantum system suffering a 
boost of momentum $\vec{k}$ (scaled magnon
liftime). If the 
characteristic relaxation time 
diverges the system does not
admit, at any finite temperature,
a fluctuation spectrum about
an ordered state. As we will later show
becomes glassy (its configurational
entropy diverges). The characteristic 
decoherence time
\begin{eqnarray}
1/(\Delta_{\vec{k}}^{(2)}E) \equiv \tau_{\vec{k}},
\end{eqnarray}
with  $\Delta_{\vec{k}}^{(2)}E $ the change
of energy of the system due to a boost of 
momentum $\vec{k}$ or scaled 
spin-wave energy. 
Whenever the integral
\begin{eqnarray}
\frac{1}{(2 \pi)^{d}} \int \frac{d^{d}k}{v(\vec{k}+\vec{q}) 
+ v(\vec{k} - \vec{q}) - 2
v(\vec{q})}
\label{MY_IN}
\end{eqnarray}
diverges then 
we may not obtain the low temperature $T=0^{+}$
magnon dispersion 
by assuming a nearly
perfectly order
state. In the above,
$\vec{q} \in M$ is any vector on 
the minimizing manifold ($M$) 
in momentum space. 
For further details the reader is referred to the Appendices.
In Appendix(\ref{M-W-I}) we generalize the Mermin-Wagner
inequality to all rotationally invariant two dimensional system 
with an analytic interaction kernel in momentum
space. In Appendix(\ref{M-W-I-boost}),
we derive inequalities in three dimensions
and interpret the extended Mermin-Wagner
theorem in terms of relaxation times.
When multiplied by the 
temperature, the integral appearing 
in the generalized Mermin-Wagner 
inequality (Eqn.(\ref{MY_IN}))
is a strict lower bound
on thermal fluctuations.
Spin systems with a non-trivial minimizing manifold 
in $\vec{k}$ space, much like spin ladders, display an 
interesting even-odd effect. When we analyze the
fluctuations of spins having an
even number of components (n),
we find precisely the integral of
Eqn.(\ref{MY_IN}) appearing 
as the relevant thermal fluctuation integral.
For a system of spins having an 
odd number of components, we find 
a more divergent (for on-shell
minima in Fourier space) fluctuation integral.
For further details the reader is referred
to \cite{me}. The bound
that we derive by a generalized 
Mermin-Wagner inequality
is a strict lower bound
on the fluctuations. When
the number of spin 
components ($n$) is
odd, the thermal fluctuations are much 
larger, and ordering
is even more inhibited.

\section{Entropy and Degeneracies}
\label{ED}

The generalized Mermin-Wagner
inequality suggests, divergent 
decoherence times are tied to a vanishing
order parameter. Physically
both effects stem from
one origin- the complicated  
energy landscape that these
systems possess. The high 
degeneracy and near degeneracy 
of low lying states makes
these systems vulnerable to thermal 
fluctuations. Metastability 
is at the heart of the 
matter. It is not surprising that
the configurational entropy
may be extensive at high
enough temperatures.

\subsection{Ground State Degeneracies}

The ground state degeneracies
of this system are obvious. As the energy
dispersion in these systems is effectively
one dimensional (it depends only on the 
radial coordinate $|\vec{k}|$), whereas 
the system is $d$ dimensional, the ground 
state degeneracy should be exponential 
in $L^{d-1}$, with $L$ the system 
size. These statements are exact
in the spherical (large $n$) limit
where every normalized superposition
of minimizing modes constitutes a 
ground state. As the number
of minimizing modes is ${\cal{O}}(L^{d-1})$
\cite{me}, the previous exponential
degeneracy follows. For special Ising
models, we can
rigorously prove \cite{me} that
the degeneracy is 
bounded from below by 
$2^{|M|}$ where $|M|$ denotes
the number of points on the minimizing manifold
which is characteristically ${\cal{O}}(L^{d-1})$.
For a similar analysis for $O(n)$ ground states
with a trivial bound $L^{n(d-1)}$ 
see \cite{me}. In the 
large $n$ limit
all of this becomes 
exact and the entropy scales as
the surface area leading to a 
{\em holographic} like
ground state degeneracy.

\subsection{Replica Calculations}

For metastable states at higher temperatures the problem 
is harder and we were not able 
to derive rigorous results.
A replica calculation within the 
self-consistent screening approximation 
reveals that the configurational entropy in all 
of these systems: continuum 
non-abelian gauge backgrounds, Kac-type
interactions \cite{loh}, Coulomb 
frustrations \cite{jorg}, and in general {\em any system} in which
the interactions are peaked on (d-1) 
dimensional shell in Fourier space, 
is {\em extensive}.

The case of Coulomb frustration
was investigated by \cite{jorg} and 
generalized to Kac-type step-function
interactions in \cite{loh}. Here we note that the minimal
ingredient in all of these calculations,
the requirement that the unperturbed
Green's function be of a simple Lorentzian
or delta-function form at low-temperatures \cite{loh,thesis}
\begin{eqnarray}
G_{0}(k) \simeq \frac{Z}{\xi^{-2}+ (\vec{k}^{2}-q^{2})^{2}},
\label{G-0-k}
\end{eqnarray}
peaked about the manifold of minimizing modes 
$|\vec{k}|=q$ (Fig.(\ref{manifold})). As in 
\cite{loh}, here $Z$ denotes the weight of the 
peak and $\xi$ the correlation length
or inverse mass gap. For any unperturbed
Green's function which may be
written as a Laurent series
in $k^{2}$, we may approximate
the Green's function in 
the vicinity of its maximum
by Eqn.(\ref{G-0-k}). 
Eqn.(\ref{G-0-k}) is the 
form analyzed long ago by Alexander and
McTague \cite{Alexander} in the context of
the liquid to solid
transition. The bare Green's function $G_{0}$
is the approximate structure
factor for
a liquid
with $l$ the average inter-atomic 
distance. If no cubic
terms favoring 
crystallization are 
present in the action
(i.e. if the transition
to the crystal
is ``avoided''),
and perform a self-consistent
screening approximation
(as done by \cite{jorg}),
then we will find a glass 
at low temperatures.
An evaluation
of the configurational entropy\cite{loh}, \cite{jorg}
\begin{eqnarray}
S_{c} = \int \frac{d^{3}k}{(2 \pi)^{3}} \{S[\frac{F}{G}]- S[\frac{F
\otimes F}{(u T)^{-1}+ G \otimes G}]\}
\label{S-C}
\end{eqnarray}
with $S[x] \equiv -x - \ln(1-x)$ shows that the entropy
is extensive within a broad temperature
range
$(T_{K} < T < T_{A} = T_{c}(q=0)/(q \pi^{2} + 1)$ \cite{jorg} and a vanishing
configurational entropy density
$S_{c} (T_{K})/V =0$ defines the Kauzmann temperature
$T_{K}$). In the above, $F$ is the Edwards-Anderson order
parameter $\lim_{t \to \infty} \langle \phi(-k,t) \phi(k,0) \rangle$,
$G$ is the standard correlation function, and $\otimes$ denotes
a convolution in Fourier space \cite{loh}. 

We now envision inverting the problem.
Instead of starting from a certain fixed model
and working our way through to find the
configurational entropy, we go back
to ask what sort of systems will exhibit 
extensive configurational entropy
(i.e. an exponentially large
number of metastable states)
by the via provided by 
calculations that invoke
the approximation of 
Eqn.(\ref{G-0-k}). 
The answer is
that the ``universality class of systems'' 
with Fourier transformed kernels
$v(\vec{k})$
having their minima
on the shells $|\vec{k}|=q>0$)
and which are analytic 
in the vicinity of these minima,
can have their Green's functions
approximated by 
Eqn.(\ref{G-0-k})
(e.g. non-Abelian background
fields, myriad translationally
invariant frustrated models
with rotational symmetry).
Once we assume this form, we
may replicate the replica calculations
of \cite{loh}, \cite{jorg} 
word for word to find that
the configurational entropy
is proportional
to the volume $S_{c}~ \alpha~ V$,
leading us to conclude that
the system displays
an exponentially large
number of metastable states.
The calculations of 
\cite{jorg},\cite{loh}
can be reproduced  
with slightly modified
expressions for
the onset temperature
$T_{A}$ of glassy dynamics
in various systems. Within all 
of these systems, including 
the dominant contribution
from the lowest eigenvalue
appearing in non-abelian gauge theories
of glasses earlier, we find
identical results. Repeating this
calculations for temperatures 
$T<T_{A}$, the self-consistency
conditions detailed in
\cite{jorg} are found to allow 
a nontrivial glassy state
wherein defects may not, in all cases,
venture more than ${\cal{O}}(q^{-1})$
at arbitrarily long times.

%For quick illustrative purposes only, we 
%now augment the $SO(4)$ system 
%of Eqn.(\ref{free-SO-4})
%by 
%\begin{eqnarray}
%\int d^{3}x ~ \sum_{M,i,n}  \frac{u_{M,i}^{n}}{4!} 
%[(\mbox{Re} \{ \phi_{M,i}^{n}(x) \})^{4} + 
% (\mbox{Im} \{ \phi_{M,i}^{n}(x) \})^{4}].
%\label{Norm-SO4}
%\end{eqnarray}
%Here, $u_{M,i}^{n}>0$, the fields $\{ \phi_{M,i}^{n}(x) \}$ the 
%Fourier transforms of the eigenmodes $\{\alpha_{M,i}^{n}(\vec{k})\}$,
%and Re, Im the real and imaginary parts. 
%The quartic term is introduced
%to force finite partition function 
%integrals. In reality, 
%the high order terms rely on the detailed microscopics.
%The reader can see that the fields
%$\{ \phi_{M,i}^{n}(x) \}$ 
%decouple and that the former scalar replica calculations
%can be repeated effortlessly by attaching
%eignevalue indices. 
For the SO(4) action 
of Eqn.(\ref{free-SO-4}),
the lowest 
rank $n=12$ representation
has $\tilde{N}=338$ 
individual real components.
Within the large $\tilde{N} \gg 1$
limit, each individual 
complex eigenvector 
has a correlation
length $\xi^{n}_{M,i}$ 
associated
with it and the configurational 
entropy 
\begin{eqnarray}
S_{c} = \sum_{M,i,n} S_{c}^{M,i,n},
\end{eqnarray}
where $S_{c}^{M,i,n}$, the 
configurational entropy associated
with each field
$\tilde{\alpha}_{M,i}^{n}(x)$ 
(the Fourier transforms of the eigenmode amplitude $\alpha_{M,i}^{n}(\vec{k})$)
is given by the self
consistent screening approximation
\cite{jorg}, \cite{loh}
for a single scalar field
(and carried in works till now
by a 
large $\tilde{N}$ calculation,
setting $\tilde{N}=1$ 
at its end).
The field associated
with the lowest eigenvalue, 
$\phi_{\min}$, has its minimizing wave-numbers 
on a finite radius, $q>0$, shell
and its configurational 
$S_{c}^{\min}$ is extensive within
the regime $T_{A}^{\min} > T > T_{K}^{\min}$.
Trivially generalizing earlier
calculations \cite{jorg}, the configurational entropy
originating from the lowest eigenvalue,
\begin{eqnarray}
S_{c}^{\min}= \frac{q_{\min}^{3}}{4 \pi}V  \Big[ 
\frac{\kappa_{\min}}{2} (1- \frac{\epsilon_{\min}}{\kappa_{\min}})^{2} 
\nonumber
\\ +
\frac{2}{\pi} \Big( (1- \frac{\epsilon_{\min}}{\kappa_{\min}})^{2} 
+ \ln(1-(1- \frac{\epsilon_{\min}}{\kappa_{\min}}) \Big) \Big].
\end{eqnarray}
Here \cite{jorg}, \cite{loh} 
$\kappa_{\min}$ satisfies 
\begin{eqnarray}
\kappa_{\min}^{2}- \epsilon_{\min}^{2} = \frac{8 \epsilon_{\min}^{2}}{\pi} 
\frac{(1- \epsilon_{\min}/\kappa_{\min})^{2}}{1-(1-\epsilon_{\min}/\kappa_{\min})^{2}} (\frac{1}{\epsilon_{\min}} - \frac{1}{\kappa_{\min}}),
\nonumber
\end{eqnarray}
with $\epsilon_{\min}^{2} = \xi_{\min}^{-2}/q_{\min}^{4}$ and $V$ the volume.
$S_{c}^{\min}(T_{c}) = C V q_{\min}^{3}$,
with the constant $C \simeq 1.18 \times 10^{-3}$. 
%The Lindemann like length \cite{jorg} $\xi_{L}^{min}$ associated
%with the lowest eigenvalue is $\xi_{L}^{\min} = 
%2/(q_{\min} \sqrt{\kappa_{\min}^{2} - \epsilon_{\min}^{2}})$.
$T_{A,K}^{\min}$ may be
for the single scalar field $\phi_{\min}$ \cite{jorg}, \cite{loh}. 
As the configurational entropy
is a sum over all eigenvalues,
$S_{c}$ is generally extensive 
over a larger range
than $T_{A,K}^{\min}$ suggest.

\section{Spatially Non-uniform Systems}

Thus far, we discussed
ideal uniform frustration. 
We now illustrate that
some of our ideas and 
calculations might not
be so restricted.
Any kernel $V_{ij} (\vec{x},\vec{y})$ 
in the general two body Hamiltonian
of Eqn.(\ref{H-g}),
%\begin{eqnarray}
%H = \frac{1}{2} \sum_{\vec{x}, \vec{y}, i,j} 
%V_{ij}
%(\vec{x},\vec{y}) \phi_{i}(\vec{x}) \phi_{j}(\vec{y}),
%\end{eqnarray}
%with $i,j$ internal field 
%components,
becomes diagonal
by some unitary transformation. The Fourier 
modes are the eigenmodes of $\hat{V}$ in
the ideal translationally invariant setting.
In general, $V_{ij}(\vec{x},\vec{y})$  
becomes diagonal in another complete orthogonal basis 
($\{ |\vec{u} \rangle \}$):
\begin{eqnarray}
\langle \vec{u}_{\alpha} | V_{i j}| 
\vec{u}_{\beta} \rangle =  \delta_{\alpha \beta}  
\langle \vec{u}_{\alpha} | V_{ij}
| \vec{u}_{\beta} \rangle \equiv  \delta_{\alpha \beta} 
v_{ij}(\vec{u}_{\alpha}) .
\end{eqnarray}
We may now examine the minimizing manifold in $\vec{u}$
space for the lowest energy eigenvector. 
Repeating calculations, 
we find that if this surface if $(d-1)$ dimensional
(e.g. a shell) and the relevant 
$v_{ij}(\vec{u})$ 
is quadratic in its
environs then generally, for large $n$, 
$T_{c} =0$. The rigorous, large $n$, 
{\em holographic} ground state degeneracies 
of the previous section and the large $n$ configurational
entropy analysis retain the same character.
In many of the expressions above,
the momentum index $\vec{k}$ 
is merely replaced by the more 
general diagonal basis index $\vec{u}$.

\section{Glassy Dynamics}

In glasses, the relaxation times
may increase by 15 orders of magnitude
over a temperature range of 100K.
It is impossible to obtain 
data below the glass transition
temperature. All data has
to be analyzed on a logarithmic
scale. As a result, it 
hard 
by principle to make precise
empirical statements in the 
vicinity of the ``glass transition''
itself (where the relaxation times,
according to common wisdom, becomes
infinite). 
We now discuss possible
non-rigorous
derivations for
both the Vogel-Fulcher (VF)
form (subsection A) 
and the avoided critical
fit (subsection B).
These two fits are 
different. There
currently is no
sufficient experimental
or mathematical
proof of any these forms.
It is, ab initio,
possible that
both forms 
and their some of 
their assumed underlying
theories are 
two aspects of the same problem.
For instance, it is possible
that the entropic-droplet/VF 
description is valid at very low temperatures 
whereas the avoided critical expression 
has its justification 
at higher temperatures
(not far from
the avoided critical
temperature itself).
Our calculations
indicate
that the 
theories of geometric
frustration in
glasses 
display both an
avoided 
critical point
and extensive
configurational 
entropy. Albeit their
suggestive character 
towards avoided critical
and/or VF dynamics,
none of these fits 
rigorously
follows from the
thermodynamics
that we investigated.

\subsection{A Possible Derivation of Vogel-Fulcher Dynamics
In Non-Abelian gauge backgrounds and others}

The basic premise of the VF fit
is that
the relaxation times
seem to diverge
at a temperature
which correlates well
with the intercept ($T_{K}$) 
of the extrapolated entropy of the supercooled 
liquid with that of the solid. As outlined in the 
previous section,
as the system is cooled down, the configurational
entropy first becomes extensive 
$T_{A}$
(at which free energy minima 
first appear);
at $T_{K}$ these minima become 
stable and the 
configurational entropy 
vanishes. 
Once we find an exponentially large number of
metastable states for $T_{A} > T > T_{K}$ 
in non-Abelian backgrounds and
other systems (by the replica 
calculations of the
previous section), we may invoke the analysis
of entropic droplets \cite{kirk}, \cite{jorg}. 
This leads to characteristic
energy barriers
\begin{eqnarray}
\Delta E  ~ \alpha ~ (TS_{c})^{-1}.
\end{eqnarray}
Linearizing the 
extensive configurational entropy,
\begin{eqnarray}
S_{c}(T \to T_{K}^{+}) \sim V (T/T_{K}-1),
\end{eqnarray}
this now leads to VF dynamics \cite{kirk}. 
In the VF fit, the relaxation
times 
\begin{eqnarray}
\tau \sim \exp[DT_{K}/(T-T_{K})]
\label{VFT}.
\end{eqnarray}
Here, $T_{K}$ is the Kauzmann temperature
at (or above) which an ``ideal glass transition'' 
would occur. At $T_{K}$ the 
{\em extrapolated} entropy of the 
liquid undergoes a crisis.

Thus, once we show that the 
lowest eigenvalue of the 
non-abelian action has a minimum 
on a shell then we may non-rigorously
derive VF dynamics. 
According
to this line of logic,
many of the translationally 
invariant systems with
minimizing momenta $|\vec{k}|=q>0$:

$\bullet$ Rigorously, do not order 
(at least in the usual
sense) at finite 
temperature (the transition
is narrowly avoided 
for $1 \gg q >0$).

$\bullet$ May display many of the characteristic thermodynamic 
signatures of narrowly avoided phase 
transition (e.g. a
crossover of the thermodynamic
functions at temperatures ``$T_{1}$''
which merge with the critical temperature
of the unfrustrated system in the limit
of zero frustration (i.e. 
when $q \to 0$)).

$\bullet$ Rigorously, have numerous degenerate
ground states, and, on a less rigorous footing, by
an insertion of Eqn.(\ref{G-0-k}) into Eqn.(\ref{S-C}),
have macroscopic configurational entropy (an exponentially
large number of metastable states).

$\bullet$ The extensive configurational entropy of 
geometrically frustrated systems
may lead to glassy 
(Vogel-Fulcher like) dynamics at low 
temperatures. We may now non-rigorously derive 
VF dynamics by fusing standard approximations \cite{kirk}
with the extensive configurational entropy 
results of the replica calculations (Section(\ref{ED})).

\subsection{A Derivation of an Avoided Critical 
Point in the Avoided Critical Dynamics
Ansatz}

As heralded in \cite{kivelson}, perhaps nature
follows another approach. One of the troubling
(or surprising) aspects of the ``glass transition''
(if it is indeed a thermodynamic transition)
is that although the dynamics undergoes a stupendous 
change- it may increase by orders of magnitude
over a small temperature window- the smooth thermodynamic
functions do not seem to know about
any impinging transition.
My coworkers and I \cite{kivelson} raised the 
possibility that the dynamics is 
governed by an avoided critical
point. This viewpoint has two 
main assumptions:

1) The system would have undergone 
a phase transition in high enough 
dimensions but due to the frustration 
encountered in low dimensional Euclidean
space it cannot. The phase transition
is narrowly avoided leading 
to the phase diagram of 
Fig.(\ref{avoid}).

2) The dynamics is tied at its 
core to the thermodynamics. The 
exponential increase in the 
relaxation times as
the ``glass transition'' 
is approached is simply
a reflection of a super-Arrhenius 
type dynamics $\tau \sim \exp (E /T)$
with the free energy barrier 
$E$ a function of the reduced 
temperature as measured relative
to the avoided critical crystallization 
temperature on the unfrustrated high dimensional
template (e.g. the 4-sphere on which 
icosahedral ordering can occur in
the case of the simple monatomic
system).

Here, we  
proved that in the large $n$ limit,
assumption (1) for the
geometrically frustrated 
actions introduced for glasses
by many \cite{book}. 
Our calculations 
yield the 
phase diagram of
Fig.(\ref{avoid})
for the $\tilde{N} \ge  338 \gg 1$ real component
theories of glasses ($n \ge 12$ in 
the $(n+1)^{2}$ complex components
of the icosahedral multipole moments
$\vec{Q}_{n}$). In Fig.(\ref{avoid}),
the ideal freezing temperature 
$T_{if} = T_{c}(1/R=0)$ is 
the transition 
temperature on 
the ideal template
that is narrowly avoided 
in flat Euclidean space. This provides 
serious conformation
to the notions underlying
the avoided critical fit. We further
proved, by a generalized Mermin-Wagner theorem,
that no standard dispersion relations 
about nearly ordered
states can exist at arbitrarily 
low temperatures for a 
host of systems, all of which display 
a minimum of the interaction kernel
at finite wave-numbers compromising
a high dimensional manifold
in $\vec{k}$ space. 
Lately, numerical results \cite{Grousson} for clock realizations
of a lattice version Hamiltonian of Eqn.(\ref{Coul}),
indeed show agreement with the form 
predicted for the scaling form
of the free energy barrier \cite{kivelson}
for this Coulomb frustrated model (Eqn.(\ref{kiv-fit})).
Furthermore, as we will now show,
if we compare the magnitude of
the avoided critical temperature
($T_{if}$) as inferred from the relaxational 
dynamics fit the free energy barrier,
we will find that it agrees very well
with what we would expect on simple
energetic grounds. We provide
further empirical
impetus for assumption (2).

\subsection{Other Relaxational Forms}

Following this work, based on a new Monte Carlo analysis, \cite{GR} 
suggested that scalar systems with an
on-shell momenta
minimum display
a pronounced slowing
down yet not nearly
as dramatic as that
in true glasses. The
dynamics is
consistent
with Hartree results. If
true, when fused with 
the contents of this
paper, this finding suggests 
that non-Abelian gauge
background trigger a novel  
sluggish dynamics
on their own right.
This new dynamics,
albeit very interesting,
will have little to 
do with actual structural glasses.
As detailed in later sections, 
structural glass theories 
must account for dynamics borne by topological
defects (ignored in the uniform
gauge background treatment).
Within the relevant 
range of investigated parameters \cite{reply},
a critical analysis
of the data of 
\cite{GR} does not 
rule out glassy dynamics.
It is possible that
none
of the forms
presented 
(including the Hartree
form) is correct
for the continuum 
models of earlier sections. 
Nevertheless, all current simulations  \cite{Grousson}, \cite{GR} 
unambiguously
point at
sluggish 
dynamics 
implying very slow dynamics
in non-Abelian backgrounds.
Ingoring topolgoical defects all
Landau-Ginzburg theories of
liquids and crystals have 
a quadratic term with $v(\vec{k})$
having minima on the shell $|\vec{k}| =q$
(augmented by cubic terms in crystals
and only quartic and higher in liquids)
\cite{Alexander}. This suggests that
if explicit topological defects are
ignored then the most
natural continuum actions for
supercooled liquids 
are those which 
we investigate.

\section{Predicting $T_{if}$ in the Avoided Critical Fit for Glasses}

We now examine the avoided critical dynamical fit
of glasses \cite{kivelson} and present a comparison 
between the average avoided critical temperatures in these
fits and the actual expected from
simple energetic considerations 
(which seem to coincide
remarkably well).

\begin{figure}
\includegraphics[width=6.5cm]{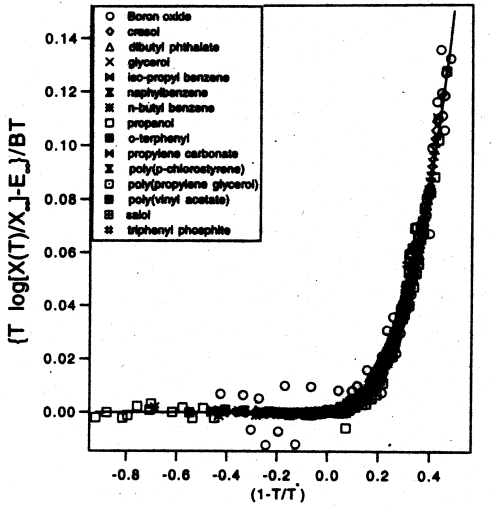}
\caption{This figure is from D. Kivelson et al. Here we fitted the 
relaxation times of all known glass formers to
the super-Arrhenius form of Eqn.(\ref{kiv-fit}). The horizontal
axis $(1-T/T^{if})$ gauges the reduced temperature
as measured relative to the transition 
temperature $T^{if}$ on the ideal curved template.}
\label{}
\end{figure}

We saw that the simplest gauge theories
of metallic glasses display an abrupt drop in their 
transition temperatures. The temperature of crystallization in 
$3-dimensional$ Euclidean space is 
depressed well below the 
crystallization temperature
on the curved ideal template.
This discontinuity might
be of profound importance:
the thermodynamics (and consequently
the dynamics) might be governed by the proximity 
to the ``ideal freezing temperature''$T^{if}$-
the temperature of crystallization
on the ideal curved space template.
In \cite{kivelson}, we demonstrated 
that the following empirical fit 
works remarkably well for all known glass formers:
\begin{eqnarray}
\tau \simeq \tau_{\infty} \exp[E(T)/(k_{B}T)]; 
E(T) = E_{\infty} + (a t)^{8/3} \theta(t) 
\label{kiv-fit}
\end{eqnarray}
Here, $t \equiv (T^{if}-T)/T^{if}$,  
the relaxation time $\tau$ is deduced from
viscosity and time dependent heat capacity 
measurements, and $\theta(t)$ is the Heavyside theta
function.

%\begin{figure}
%\includegraphics[width=9.7cm]{fig10aa.eps}
%\caption{This figure is from D. Kivelson et al.}
%\label{}
%\end{figure}

%\begin{figure}
%\includegraphics[width=9.4cm]{fig11aa.eps}
%\caption{Comparison between the Kivelson fit and the 
%widespread Vogel-Fulcher fit. From D. Kivelson et al.}
%\label{}
%\end{figure}

We may expect $k_{B}T^{if}$ to be indicative of the 
cohesive energy of the ideal curved space crystal, and 
the natural melting temperature scale ($k_{B}T_{melt}$) 
to be correlated with
the crystallization 
energy of the real (space) structure.
Obviously entropic effects are important
(e.g. those present in the BCC $\rightarrow$ FCC transition \cite{Alexander}). Amongst these liquids, the average value of 
$x  \equiv [T^{if}-T_{m}]/T_{m}$
is $9.59\%$. We may 
compare this to the 
value computed by Frank \cite{Fr} for the relative
cohesive energy, $(E^{icosahderal} - E^{FCC})/E^{FCC}$, 
for a simple Lennard Jones
liquid, which leads
us to expect  $\langle x \rangle =8.4\%$. 
The proximity of these two values is misleading.
$x$ varies dramatically from one liquid to another-
the ideal crystalline structures, and their respective energies
change radically from one material to another. 
Cautious optimism is still called for.
If the theory were reliable and the parameter $T^{if}$ obtained
from the fit could indeed be interpreted as the freezing temperature on the
ideal template, one would expect that, for all liquids,
$x>0$, as the structure on the ideal template (where the ideal 
minimum energy local structure could be extended in all
directions) is by its very definition
more stable than (or at least as stable as) the 
corresponding structure constrained to reside
in Euclidean space.

\begin{tabular}{|r|l|}
\hline \hline
glass former & $x \equiv [T^{if}-T_{m}]/T_{m}$ \\
\hline
n-butyl benzene & 7.02\% \\
triphenyl phosphite & 8.2 \% \\
isopropyl benzene & 20.6 \% \\
propylene carbonate & 7.73 \% \\
salol & -4.4 \% \\
dibutyl phthalate & 21 \% \\
o-terphenyl & 5.7 \% \\
s-trinaphthyl benzene & 9.1 \% \\
n-propanol & 30.6\% \\
$\alpha$-phenyl-cresol & -10.3 \% \\
glycerol & 9.9 \%
\\
\hline \hline
glass former average &  9.59 \% \\
\hline \hline
simple theory  & 8.4\% \\
\hline
\end{tabular}

\section{Avoided Critical Dynamics And Microscopics}

Ideal order cannot tessellate Euclidean space
and it may do so only on a higher dimensional 
surface with energetics
that seem to concur with the 
picture that we have in mind.
In order to make progress toward understanding
the dynamics we need to explicitly see
how things transpire in real Euclidean
space. We may adopt two trivial points of
view: either the action is
pinned to be that of a system
in a {\em fixed non-Abelian}
background, or 
the distribution of
matter alters to 
the geometry by virtue of
the topological defects that
it carries in which 
case we have a full
blown {\em matter coupled
gauge theory}- 
an elastic theory
with disclinations
where the disclinations
are not pinned down.
The ideal 
order on the curved high dimensional
template needs to be mapped
into flat space by introducing
the disclinations of Rivier
to compensate for the
lack of curvature in Euclidean
space.

\subsection{Fluctuating Geometry}

Let us assume that the ideal curved order
is neutralized in flat Euclidean space
by the insertion of a uniform disclination
density (giving rise to an effective
negative curvature to compensate for the 
positive curvature on the torn pieces
of the ideal sphere). When glued together
the curved pieces of the fractured sphere form 
disclinations at their common junctions.
This is much like the seams that we 
encounter in orange peals that need
to glued together somehow to form a
flat surface. The disclination
lines can easily be traced
by looking at the locus of
all points that have a lower
or higher than usual coordination
number as compared to that present in
the ideal icosahedral tessellation.
Formally, they can be seen by decurving
processes relating the curved template
to flat space. To a zeroth order approximation,
the additional energy of the decurved structure 
(an elastic strain energy) amounts to the total
length of the disclination network. In the 
context of decurving the \{3,3,5\} polytope,
the net number of disclinations
is dictated by the number of sites having
a coordination number different from
the ideal (12-fold coordinated
vertices). An analysis along
this counting may be found in
\cite{book}. If we let 
the field $\phi$ denote the
density of disclination
lines then in 
our analysis such a
zeroth order effect amounts
to a chemical potential term
$\mu \phi$ with the chemical 
potential $\mu$ now denoting the energy per unit
disclination line.

Disclinations (or disclination pairs-
dislocations) interact by direction
dependent Coulombic like forces. 
In a two dimensional XY model like 
setting (e.g. vortices in a thin film 
superconductor) where the disclinations
amount to additional twists that the 
phase must undergo when 
parallel transported 
about their cores. Such 
fractional vortices interact
Coulombically. Two fractional vortices
of opposite charges  
undo the damage that the other 
does at long distances. In three
dimensions, the energy would 
be minimal for such wedge disclinations
(or fractional vortices) lying in
the same plane. 
In the tensorial elastic setting,
such angle dependent forces undergo
only simple detailed change. The Blin interaction
between dislocations is a decorated
Biot Savart interaction \cite{kleinert}.

The same occurs in gravity. In 2+1 dimensional 
gravity, masses amount to conical defects which cut
and remove (or insert) additional angles in the plane 
much like disclinations.
If we disregard ``unimportant'' details,
trying to get to the heart of the matter
by noting that the defect densities interact
Coulombically, then we may immediately
provide a possible derivation of the 
Coulomb frustrated Hamiltonian 
of Eqn.(\ref{Coul}). 
We now try to give the reader
a quick intuitive grasp of
the microscopic derivation
that what lacking
till now (as still is).
Let $\phi$ denote the defect
density. In reality $\phi$ is
a tensor by let's disregard details and
merely see how things may work.
For simplicity, we set the field $\phi$ be the local 
scalar curvature. It is positive inside 
the broken fragments of 
the ideal sphere that have been flattened out;
the net curvature is superposed on a neutralizing
background that ensures that the net curvature 
is zero (as in the Frank-Kasper phases). Once we make the correspondence 
among the ideal sphere fragment densities
and the disclination densities,
eqn.(\ref{Coul}) may, trivially,
follow. The disclinations (or 
disclination pairs) interact Coulombically
and the density toward ideal (high dimensional 
spherical) order within each resulting spherical 
fragment is captured by the gradient term. 
These two terms compete
leading to extensive entropy
at low temperatures
and glassy dynamics. However,
the dynamics is not 
trivially of a 
Vogel-Fulcher form.

To give the discussion a 
more detailed flavor we 
now give some 
of the standard 
expressions.
Denoting by $\vec{u}$ the displacements
of atoms from their ideal equilibrium positions, 
the dislocation density reads
\begin{eqnarray}
\alpha_{ij} = \epsilon_{ikl} \partial_{k} \partial_{l} u_{j}.
\label{alph}
\end{eqnarray}
In a gravitational language, the dislocation 
density is, in simple disguise, the connection 
determining the change the of a vector due to 
parallel transport alone,
\begin{eqnarray}
\alpha_{ij} = \epsilon_{ikl} \Gamma_{klj} 
\end{eqnarray}
with the connections
\begin{eqnarray}
\Gamma_{ijk} = \partial_{i} \partial_{j} u_{k}. 
\end{eqnarray}
The connection between disclinations and
gravitational curvature is even more transparent. With
\begin{eqnarray}
\omega_{ij} = \frac{1}{2}(\partial_{i} u_{j} - \partial_{j}
u_{i}),
\end{eqnarray}
the disclination density 
is 
\begin{eqnarray}
\Theta_{ij} = \frac{1}{2} \epsilon_{ikl} \epsilon_{jqs} \partial_{k}
\partial_{l} \omega_{qs},
\end{eqnarray}
while the curvature tensor is
\begin{eqnarray}
R_{ijkl} = (\partial_{i} \partial_{j} - \partial_{j}
\partial_{i}) \omega_{kl}.
\end{eqnarray}
An explicit calculation for a wedge disclination 
corresponding to the insertion of an additional
angle (of magnitude $\alpha$) in between the 
lips of a cut in the 12 plane shows
that
\begin{eqnarray}
R_{1212} = - \alpha \delta^{2}(x^{\lambda})
\end{eqnarray}
where $\delta^{2}(x^{\lambda})$ denotes the delta function
in the plane for the disclination line $x^{\lambda}$
threading three dimensional space. 
Thus, the curvature tracks the 
disclination lines and its magnitude
is proportional to the 
disclination angle. The disclination
lines of Rivier \cite{rivier} become mass sources.
For a positive inserted angle,
the scalar curvature is negative. 
If we wish to squash the angles to
close the gaps such as those
appearing when trying to
piece together five tetrahedrons
around a common edge.
we insert a negative angle.
The curvature, the full contraction of the curvature
tensor,
\begin{eqnarray}
R= R_{\mu \nu}^{\mu \nu} 
\end{eqnarray}
is positive on a sphere (scaling
as the inverse squared radius).
The Blin interaction between dislocations 
\begin{eqnarray}
E = \frac{\mu}{T} \int d^{3}x ~d^{3}x^{\prime} ~
\alpha_{ij}(x) [K] \alpha_{l^{\prime} j^{\prime}}, 
\end{eqnarray}
with the distance dependent part of the kernel $[K]$
\begin{eqnarray}
- (\frac{\epsilon_{l
l^{\prime}k}  \epsilon_{j
j^{\prime}k}}{r})  + \frac{1}{1- \nu} \epsilon_{l jk}
\epsilon_{l^{\prime} j^{\prime} k^{\prime}} \partial_{k}
\partial_{k^{\prime}} r,
\end{eqnarray}
where $\mu$ and $\nu$ are the elastic constants, and $\alpha$ 
the dislocation densities of Eqn.(\ref{alph}). All interactions
between elastic defects are of a Coulombic 
character which has been tensorially flavored 
up. We can easily see why this must be so: 
Elasticity is centered around a second order gradient
expansion,
\begin{eqnarray}
{\cal{L}} = \frac{1}{2} c_{ijkl} u_{ij} u_{kl}, \nonumber
\\ u_{ij} \simeq \frac{1}{2} (\partial_{i} u_{j} + \partial_{j} u_{i}) ,
\end{eqnarray}
with $u$ the displacements about an equilibrium position. 
When Fourier transformed
and inverted, the Green's functions through which 
the topological defects communicate always 
give rise to $k^{-2} \sim r^{2-d}$ interactions. 
In two dimensions, the interaction between
the dislocations is, not too surprisingly, 
logarithmic. As all derivations that
we have been able to
devise (apart from
the generalized Mermin-Wagner
theorem and the construction
of $O(n)$ ground states) have 
been approximate, we examine 
numerical results. Lately, 
computations \cite{Grousson} for clock realizations
of a lattice version Hamiltonian of Eqn.(\ref{Coul}),
indeed show agreement with the form 
predicted for the scaling form
of the free energy barrier \cite{kivelson}
for this Coulomb model (Eqn.(\ref{kiv-fit})).
We have given simple arguments for the form of
this fit, although a real theory is still 
lacking.  As noted, the geometrical
frustration breaks the glass into  
clusters. In an earlier publication
\cite{kivelson}, we have found that assuming 
the usual Ornstein-Zernicke form for $T<T_{c}$,
within our ``standard model'',
cumulant crossing occurs 
when the size of the domains is
larger than $R_{D} \simeq \xi_{0}^{-1}$,
with $\xi_{0}$ the correlation
length of the unfrustrated
system which diverges
at the (avoided) critical
temperature $T_{c}$. A somewhat similar result
(with $\xi_{0}$ replaced 
by the correlation length $\xi$
of the frustrated system) followed from
our 1/n calculation \cite{joe}
for the Josephson like correlation
length. This 1/n calculation result
undergoes no change for
any of the models that 
have their free energy minima 
on a shell in Fourier 
space which becomes 
sharp at 
low temperatures.

To shortly examine viable
extensions later on, 
we sketch our derivation in \cite{kivelson}.
Usually, the specific heat exponent
$\alpha$ is small 
and consequently from the
hyper-scaling relation:
\begin{eqnarray}
\nu d = 2 - \alpha,
\end{eqnarray}
we expect $\nu \simeq 2/d$.
The free energy barriers that govern the low temperature
dynamics involve the flipping of domains 
(of size $R_{D}$). We assume
that all is governed by the relative 
proximity to the avoided
critical point. As in Eqn.(\ref{kiv-fit}), we
consider temperatures below the avoided critical
temperature $T< T^{if}$ and denote by
$t$ the absolute value of the 
relative reduced temperature.
The correlation length
in the unfrustrated system
$\xi_{0} \sim t^{-\nu}$.
The surface tension in
the unfrustrated system
\begin{eqnarray}
\sigma \simeq B \frac{T_{if}}{\xi_{0}^{d-1}},
\end{eqnarray}
with $B$ a constant. The free energy barriers 
that we might find by breaking
the medium into unfrustrated patches
of size $R_{D}$ (the geometrical
frustration in this 
approximation merely
breaks the unfrustrated medium
into finite patches (domains) 
of scale $R_{D}$), 
\begin{eqnarray}
E \simeq \sigma R_{D}^{d-1} \simeq B T_{if} (\frac{R_{D}}{\xi_{0}})^{d-1}
\end{eqnarray} 
Employing our estimates \cite{joe,kivelson} $R_{D} \sim \xi_{0}^{-1}$ and
$\nu \sim 2/d$,
\begin{eqnarray}
E \sim t^{4(d-1)/d}.
\end{eqnarray}
Substituting $d=3$ this gives an exponent
of $8/3$ which which works best in our
fit. It should be noted, however, that 
if we assume that 
\begin{eqnarray}
E(T) =  E_{\infty} + (at)^{x}
\end{eqnarray}
then for we can obtain
reasonable fits with 
$7/3 \le x \le 3$.

\subsection{Fixed Non-Abelian Background}

Here no derivation of Eqn.(\ref{kiv-fit})
seems possible along the lines
of the previous derivation.
We may read off the dispersion 
from the lowest eigenvalue.
For instance, in our
toy model for the 
SO(3) action, $\lambda_{-}(k) = \mu+ k^{2}+ 3 \theta^{2}/2- \frac{1}{2}
\sqrt{16 k^{2} \theta^{2} + \theta^{4}}$
is the relevant dispersion. For $k \gg \theta$, (distances 
small compared to the zero temperature modulation length- 
the radius of the ideal sphere) the square
root may be expanded and to lowest
order, the dispersion is trivially
$(k^{2}-2 k \theta + \mu)$. In a 
spin model, such a dispersion 
corresponds to a short
range nearest neighbor
ferromagnetic interaction ($k^{2}$)
frustrated by a longer range anti-ferromagnetic
interaction $(- 2 \theta |\vec{k}|)$.
Unfortunately, the frustrating interaction
is not of a much longer
algebraic decay than the 
unperturbed ferromagnetic
interaction. In a cumulant 
expansion similar to
the one that
we performed in \cite{kivelson},
we now find
that the characteristic 
domain size is $R_{D} \sim \xi_{0}^{2}$.
When inserted back into our
derivation for the free energy 
barriers this gives physically
impossible dynamics: $E \sim t^{2(1-d)/d}$,
a free energy barrier that quickly diminishes
as we go away from the avoided critical
point. For $k \ll \theta$ this
worsens. We cannot see a natural
generalization of the frustrated domains
of the previous section which
provides sensible results.

To summarize, for non-Abelian gauge backgrounds,
we can very easily derive Vogel-Fulcher
dynamics yet we are unable to derive
avoided critical dynamics of the 
form of Eqn.(\ref{kiv-fit}). Notwithstanding,
these models do display avoided 
critical behavior insofar as their 
thermodynamics is concerned. Furthermore,
at low temperatures only the Fourier
modes close to the minimizing
shell should be of any importance.
If a low temperature fit of the
form of Eqn.(\ref{kiv-fit})
works well for the Coulomb frustrated Ferromagnet
at small yet finite frustrations $e$, then 
Eqn.(\ref{kiv-fit} might also capture
the low temperature dynamics of all models
(including the fixed non-Abelian background)
that have their minima on a shell of 
finite radius in $k$ space. The approximate 
identification can be made by Taylor expanding the 
interaction kernels in the various 
cases to quadratic order and seeing that
they all match.

\section{Conclusions}

To summarize, we have shown
that a multitude of translationally
invariant systems having their
minimizing modes lie on a (d-1) 
dimensional shell in $\vec{k}$ space

{\bf 1})  Do not allow
for a magnon dispersion
relation about an ordered
state at any finite temperature
whenever a certain
(Mermin-Wagner like) integral is 
seen to diverge.

{\bf 2}) We have noted that the generalized Mermin-Wagner
inequality relates the magnetization (or its absence
whenever entropy dominates)
and the average of the expected 
relaxation times within the system.

{\bf 3}) The correlation and modulation
lengths generally exhibit dynamics as 
a function of temperature
(the imaginary time width).
Crossovers in the asymptotic
form of the correlation
functions do, in general, 
occur at a finite temperature
which veers toward the critical temperature
of the unfrustrated system in the limit
of zero frustration (zero  
curvature in myriad geometric models).
 
{\bf 4}) The systems exhibit an astonishing
ground state and near ground state 
degeneracy. We may easily provide
rigorous lower bounds on the ground
state degeneracies and
show, via replica calculations,
that the configurational entropy is
extensive. Physically, it is this
multitude of degenerate and nearly degenerate
states that makes the system so fragile to
thermal perturbations. As a consequence,
the ordering temperature the ordering temperature
is very low.

{\bf 5}) We discussed how in a multitude of 
a systems (Non-Abelian backgrounds and
all others with an ``on-shell'' minima 
in Fourier space), the derivable
extensive configurational entropy
suggests Vogel-Fulcher like 
dynamics.

{\bf 6}) We illustrated how many
of these notions can be extended
to non-uniform system 
whereby the interactions
are diagonal in 
a non-Fourier basis.

{\bf 7})  We have shown how the dynamics may indeed
correlate with the avoided critical fit of
\cite{kivelson} by comparing the melting
temperatures with the fitted temperatures.
The average values are in good agreement
with our theory.

{\bf 8}) We suggested a via for
strengthening the link between  
avoided critical dynamics \cite{kivelson}
with microscopic consideration regarding
interactions between disclinations.

\begin{acknowledgments}

Most of this work was covered in
my thesis long ago \cite{thesis}. During my 
thesis years I had the good fortune of working 
with many good friends. 
A few deserve close mention-
S. A. Kivelson, 
J. Rudnick, D. Kivelson, G. Tarjus, and 
L. Chayes. I also wish to 
thank D. N\'ogr\'adi and J. Sethna
for a discussion.

\end{acknowledgments}

\appendix

\section{$n \to \infty$ Avoided Criticality In The quantum Arena}
\label{qu_n}

The single  $n \rightarrow \infty$  parquet diagram (Fig.(\ref{largen}))
reads 
\begin{eqnarray}
\Sigma^{0} = \frac{nu T}{2} \sum_{\omega_{m}}\int \frac{d^{d}k}{(2 \pi)^{d}}~~
 \frac{1}{\omega_{m}^{2}+ v(\vec{k}) + \mu_{min}},
\end{eqnarray}
with the bosonic Matsubara frequencies $\omega_{m} = 2 \pi mT$.
\begin{eqnarray}
\Sigma^{0} = \frac{nu T}{2} ( \frac{1}{2 \pi T})^{2} \sum_{m} \int
 \frac{d^{d}k}{(2 \pi)^{d}} [m^{2}+( \omega_{\vec{k}}/(2 \pi T))^{2}]^{-1}
\end{eqnarray}
where $\omega_{\vec{k}}= \sqrt{\mu+ v(\vec{k})}$.
The sum can be evaluated by the method of residues which gives
\begin{eqnarray}
\sum_{m=-\infty}^{\infty} [m^{2}+ y^{2}]^{-1} = \frac{\pi}{y} \coth \pi y
 \mbox{ ~ for $y>0$}.
\end{eqnarray}
Using this we obtain 
\begin{eqnarray}
\Sigma^{0}= \frac{1}{4} \int \frac{d^{d}k}{(2 \pi)^{d}}
 \frac{1}{\omega_{\vec{k}}} \coth (\frac{\omega_{\vec{k}}}{2 T}).
\end{eqnarray}
From the identity 
\begin{eqnarray}
\coth (\frac{x}{T}) = 1+ 2 n_{B}(\frac{2x}{T})
\end{eqnarray}
where $n_{B}(\frac{x}{T})$ represents the bosonic distribution function
\begin{eqnarray}
n_{B}(\frac{x}{T}) = [\exp(\frac{x}{T})-1]^{-1}.
\end{eqnarray}
Thus 
\begin{eqnarray}
\Sigma^{0} = \frac{nu}{4} \int \frac{d^{d}k}{(2 \pi)^{d}} 
 \frac{2n_{B}(\frac{\sqrt{v(\vec{k})+\mu}}{T})+1}{\sqrt{v(\vec{k})+\mu}}.
\end{eqnarray}
As $\mu \rightarrow \mu_{\min} = -v(\vec{q})$, this integral diverges
more strongly than in the 
classical case. This divergence 
remains down to zero temperature.
The integrand 
is (up to a constant) the static correlation function.
\begin{eqnarray}
G(\vec{k}) = \frac{1}{2}
\frac{2n_{B}(\frac{\sqrt{v(\vec{k})+\mu}}{T})+1}{\sqrt{v(\vec{k})+\mu}}.
\end{eqnarray}

\section{The SO(4) soccer-ball model on a cubic lattice}
\label{SO_4_SOCCER}

For $O(4) \mbox{ spins with }SO(4)$ couplings in $d=3$:
\begin{equation}
  D_{\nu}S^{\mu} = \partial_{\nu}S^{\mu} + S^{\lambda}
  \Gamma^{\mu}_{\nu \lambda}.
\end{equation}

For four-component spins lying on a three-dimensional lattice 
(i.e. when the continuum limit is not taken) having $SO(4)$
 couplings ``$[-\vec{S}(\vec{x})*\vec{S}(\vec{y})]$''
the matrix $-v_{i,j}(\vec{k})/2$ 
reads 
$$
\pmatrix{ A & 0  & 0 & -E \cr
 0 &  B  & 0 &  -F \cr
0 & 0 &  C & -G \cr
E  & F  &  G  & D }.
$$
where
\begin{eqnarray}
A= \cos \theta \cos k_{1}+ (\cos k_{2}+ \cos k_{3}) \nonumber
\\ B= (\cos k_{1} +\cos k_{3}) +\cos \theta \cos k_{2} \nonumber
\\ C= \sum_{l=1}^{2} \cos k_{l}+ \cos \theta \cos k_{3} \nonumber
\\ D=  \cos \theta \sum_{l=1}^{3} \cos k_{l} \nonumber
\\ E= i \sin \theta \sin k_{1} \nonumber
\\ F= i \sin \theta \sin k_{2} \nonumber
\\ G=  i \sin \theta \sin k_{3}
\end{eqnarray}

Note the appearance of a non-uniform
sign (and nontrivial phases)
linking the various spin 
components. These play
the role of frustrating 
interactions present
in scalar theories;
in a high temperature
expansion for the partition
function $Z$, not all
terms are the sum
of different contributions,
all adding in unison with
the same phase (and therefore
also sign). The alternating
signs of the resulting
constituent terms are
analogous to those generated
in other frustrated systems
by competing ferromagnetic
and antiferromagnetic
interactions in the evaluation
of the partition 
function $Z$.  For a nonzero value of the angle $\theta$,
~$T_{c}(\theta \neq 0)=0$. The minimizing modes
lie on a $(d-1)$ dimensional shell.

\section{Explicit Expressions for correlation function crossovers}
\label{explicit_crossover}

\subsection{Uniform SO(3) and SO(4) backgrounds}

For the $SO(3)$ action of \cite{subir} in d=2 dimensions, summing over a 
set of $(2\ell+1) \times
(2\ell+1)$ representations with corresponding rank spherical tensor
order parameters:
\begin{equation}
  H = \frac{1}{2} \sum_{\ell} \int d^{2}x {[K_{\ell}|(\partial_{\mu}+\imath
    \theta \epsilon_{\mu \nu}
    \hat{L}^{(\ell)}_{\nu})\vec{\phi}_\ell|^{2}+\mu_{\ell}|\vec{\phi_{\ell}|^{2}}}],
\end{equation}
with $\{ \mu_{ell} \}$ chemical potentials- Lagrange multipliers- set
to secure overall normalization of $\{ \vec{\phi}_{\ell} \}$.
In $\vec{k}$-space,
\begin{eqnarray}
  H =\frac{1}{2N} \sum_{\ell} \sum_{i}
  \sum_{\vec{k}}|\alpha_{\ell}^{i}(\vec{k})|^{2} [K_{\ell}
  \lambda_{\ell}^{i}(\vec{k})+\mu_{\ell}]; \nonumber
\\ \mbox { with } \phi_{\ell
    m}(\vec{k}) = \sum_{i=1}^{2 \ell+1}\alpha_{\ell}^{i}(\vec{k})
  e_{\ell m}^{i}(\vec{k}).
\end{eqnarray}
The self-consistent equation of constraint that the Lagrange 
multipliers must satisfy is
\begin{equation}
  \frac{1}{k_{B}T}= \int \frac{d^{2}k}{(2\pi)^{2}}G(\vec{k}).
\end{equation}

The inverse critical temperature, $\frac{1}{k_{B}T_{c}}$, diverges as the reciprocal of a lower cutoff on
$|k-q|$.  We find that ($2\ell+1$) poles are generically present for each $\ell$
when Fourier transforming
\begin{equation}
  G(\vec{k}) = \sum_{\ell} \frac{2 \ell +1}{ 4 \pi} \sum_{i} \langle
  \alpha_{\ell}^{i}(\vec{k}) \alpha_{\ell}^{j}(- \vec{k}) \rangle~ e_{ \ell
    0}^{i}(\vec{k}) e_{\ell 0}^{j}(-\vec{k}),
\end{equation}
which simplifies to
\begin{equation}
  G(\vec{k}) = \sum_{\ell} \frac{2\ell+1}{4 \pi} \sum_{i=1}^{2\ell+1}
  \frac{k_{B}T}{K_{\ell} \lambda_{\ell}^{i}(\vec{k})+\mu_{\ell}}
  |e^{i}_{\ell,0}(\vec {k})|^{2}.
\end{equation} 
Our two-dimensional soccer-ball ($\ell=1$) Hamiltonian: 
\begin{eqnarray}
H=
-\sum_{ \langle \vec{x},\vec{y}
\rangle}\vec{S}(\vec{x})*\vec{S}(\vec{y})
\end{eqnarray} 
leads to three poles at all temperatures. Much unlike the
pole locations, the branch points are temperature independent. 
If we ignore the $\vec{k}$ dependence of
$|e_{1,0}^{\pm}(\vec{k})|^{2}$, then the contribution to $G(|\vec{x}|)$ from 
the branch cut contour will read
\begin{equation}
\frac{\theta}{4 \pi x^{2}} (A_{+}-A_{-}) \int_{\theta/4}^{\infty}
dv~\frac{ v ~\sqrt{\theta^{2}-16v^{2}/x^{2}}~I_{0}(v)}{(\mu-\frac{v^{2}}{x^{2}}
+\frac{3}{2}\theta^{2})^{2}
 -\theta^{2}(\theta^{2}-16\frac{v^{2}}{x^{2}})},
\end{equation}
where $A_{\pm} = \frac{3}{4 \pi} |e_{1,0}^{\pm}|^{2} k_{B}T$
and $I_{0}$ a Bessel function
of imaginary argument. 
The contributions from the branch cut
would seem (if $\{ e_{\ell=1,0}^{\pm} \}$ were $\vec{k}$ independent)
to give rise to quasi-algebraically damped correlations.

For the case at hand,
\begin{eqnarray}
|u_{-} \rangle = N_{-} | \frac{4 i \theta k_{y}}{\theta^{2}-\sqrt{16 k^{2} \theta^{2}
 + \theta^{4}}},
 \frac{ 4 i \theta k_{x}}{\theta^{2}- \sqrt{16 k^{2} \theta^{2}
+\theta^{4}}},- 1 \rangle,
 \nonumber
\\ |u_{+}\rangle = N_{+} |\frac{4 i \theta k_{y}}{\theta^{2}+\sqrt{16 k^{2} \theta^{2}
 +\theta^{4}}},
 \frac{4 i \theta k_{x}}{\theta^{2}+ \sqrt{16 k^{2} \theta^{2}
+\theta^{4}}},1 \rangle, \nonumber
\end{eqnarray}
are the eigenvectors expressed in the Cartesian basis. 
In the spherical tensor basis $\{|e_{\pm}^{1,0}|^{2}\}$ 
are the squared moduli $|N_{\pm}|^{2}$,
of the z-components of the eigenvectors $|u_{\pm} \rangle$
are
\begin{eqnarray}
|e_{\ell=1,0}^{\pm}|^{2} = 1 - \frac{8 k^{2} \theta^{2}}{\theta^{4}
 \pm \theta^{3} \sqrt{16 k^{2} + \theta^{2}} 
+ 16 \theta^{2} k^{2}}.
\end{eqnarray}
The dependence of $|e_{\ell=1,0}^{\pm}|^{2}$ on $|\vec{k}|$ is manifest.
Inserting this in $G(\vec{k})$ and Fourier transforming
one readily verifies that no quasi-algebraic behavior remains
only the usual exponential correlations originating from 
simple poles appear. For infinitesimal curvature (or frustrations)
$\theta$, 
all correlation lengths (originating from these poles
 $|Im\{k_{o}\}|^{-1},|Im\{k_{\pm}\}|^{-1}$ can be 
made as large as desired when $T \rightarrow 0$: $k_{o}^{2} = 
- [\mu+\theta^{2}],
 ~ k_{\pm}^{2} = \frac{1}{2} [ \theta^{2} - 
2 \mu \pm i \sqrt { 16 \mu \theta^{2} + 7 \theta^{4}}] $.
Note that here there is no crossover of a correlation length
into a modulation length: The poles $\{\pm k_{\pm}^{2}\}$ are
complex conjugate for  $\mu>\mu_{\min} = -7 \theta^{2}/16$
and will give to a damped oscillatory behavior at all
temperatures. Nonetheless, a crossover does occur
(albeit not a thermodynamic one). At very low temperatures
the long correlations are spawned by $k_{\pm}$ and are 
are therefore oscillatory.  At intermediate temperatures the 
correlation length generated by $k_{0}$ can be larger leading to
asymptotic long range correlations that
are non-oscillatory in character. To see this, we note that 
\begin{eqnarray}
|Im\{k_{0}\}| = \sqrt{\mu} \nonumber
\\ |Im\{k_{\pm}\}| = (2 \theta^{4}+ 3 \mu \theta^{2}
+ \mu^{2})^{1/4} |\sin(\phi/2)| \nonumber
\\ \phi = \tan^{-1}  (\frac{\sqrt{16 \mu \theta^{2} + 
\theta^{4}}}{\theta^{2}-2 \mu}).
\end{eqnarray}
Whenever $\mu(T) = \theta^{2}/2$,
\begin{eqnarray}
|Im \{ k_{0} \}| = \frac{\theta}{\sqrt{2}}, \nonumber
\\ |Im \{ k_{\pm} \}| = (3.75)^{1/4} |Im \{ k_{0} \}|.
\end{eqnarray}
At this temperature $| Im \{ k_{\pm} \} | > |Im \{ k_{0} \}|$
and therefore the asymptotic correlations will not be oscillatory
as they are near the ground state.   
In this way we can identify a low temperature 
crossover. It is straightforward to obtain a
simple expression
for the value of the chemical potential 
$\mu$ at this crossover temperature.

\begin{eqnarray}
|Im\{k_{\pm}\}| = 2^{-1/2} (2 \theta^{4}+ 3 \mu \theta^{2}
 + \mu^{2})^{1/4} \nonumber
\\ \times \
\sqrt{1-[1+\frac{16 \mu \theta^{2}+ \theta^{4}}{(\theta^{2}
-2 \mu)^{2}}]^{-1/2}} 
\nonumber \\
= |Im \{k_{0} \}|= \sqrt{\mu}.
\end{eqnarray}

For larger matrices (e.g. $\ell>1$ in an $SO(3)$ action)
there are potentially more crossovers. 
For these matrices, at asymptotically large temperatures all
correlation lengths $\rightarrow \mu^{-1/2}$.
As we will see, this is not the case for long range interactions
Notice that this change from a uniform behavior at 
high $T$ to finite modulation lengths at low $T$ 
is a feature shared in common with the Coulomb 
frustrated ferromagnet (and any other system
with a frustrating long range interaction).
We find similar results for the $SO(4)$ action in three dimensions of 
\cite{subir}. The correlator 
\begin{equation}
  G(\vec{k})= \sum_{n} \frac{n+1}{2 \pi^{2}} \frac{k_{B}T}{K_{n}
    \lambda_{p=0,i}^{n}(|\vec{k}|)+\mu_{n}} |\sum_{m}
  e_{n,mm}^{i}(\vec{k})|^{2},
\end{equation}
with $i$ a single index labeling the eigenvalues, where we
note the commutativity of $[M_{+}-M_{-}]$ with the kernel.

\subsection{Correlation Function Crossovers Within the Coulomb
Frustrated Ferromagnet}

Unlike other Appendices, most of the contents of
this section appear in a Letter version of
our work on Coulomb frustration \cite{joe}. 
%Here 
%we re-introduce these so that
%the reader may appreciate discerpencies
%between the multiple correlation lengths
%appearing in
%short range  
%systems to follow in the
%next appendices (where the number
%of correlation lengths
%is constant) and
%longer range interactions
%investigated here where the number of
%correlation
%lengths may vary 
%as a function of
%temperature.
The continuum limit Hamiltonian of the Coulomb frustrated 
ferromagnet
\begin{eqnarray}
  H = \frac{1}{2} \int \frac{d^{d}k}{(2 \pi)^{d}} (k^{2}+ e^{2}k^{-2} + \mu) 
| S(\vec{k})|^{2}
\end{eqnarray}
where the spherical constraint (global spin normalization) is 
enforced
via the chemical potential $\mu$. A non-increasing weight
for all modes
$\vec{k}$ mandates $\mu \geq -2e$. The
minimizing modes 
$\vec{q}^{2} = e$).  By equipartition, the mode occupancy 
$\langle |S(\vec{k})|^{2} \rangle
= k_{B}T/(k^{2}+e^{2}/k^{2}+\mu)$. 
The correlator:
\begin{eqnarray}
G({\vec  x}) = && \frac{1}{(2 \pi)^{3}} \int d^{3}k 
~\langle|S({\vec k})|^{2}\rangle~ 
\exp[i {\vec k} \cdot {\vec x}] \nonumber 
\\ = && \frac{k_{B}T}{2 \pi^{2}|{\vec  x}|}\int_{0}^{\infty} dk
  \frac{k^{3}
    [Im\{{e^{ik|{\vec  x}|}\}]}}{(k^{2}+\alpha^{2})(k^{2}+\beta^{2})}
\end{eqnarray}
where
\begin{equation}
  \alpha^{2}, \beta^{2} = \frac{\mu \mp \sqrt{\mu^{2}-4e^{2}}}{2}.
\end{equation}
When $\mu > 2 e$, the integral can be readily evaluated
by applying the residue theorem to the poles lying on the
imaginary axis at $k=\pm \imath \alpha ,\pm \imath \beta$, 
\begin{eqnarray}
  G({\vec  x}) = && \frac{k_{B}T ( \beta^{2} e^{-\beta |{\vec  x}|} 
  - \alpha^{2} e^{-\alpha |{\vec  x}|} ) }{4 \pi |{\vec  x}| 
  (\beta^{2}-\alpha^{2})}. 
%\nonumber \\.
\label{eq:g}
\end{eqnarray}
Note the existence of two macroscopic correlation lengths -- 
a consequence of charge neutrality.
As the spins portray charges, they 
must sum to zero,
\begin{equation}
\int G({\vec  x})~d^{3}x= \langle |\int S({\vec  x})~d^{3}x|^{2} \rangle =0.
\end{equation}
Whenever $G$ is dominated by its long-distance behavior, the integral 
can vanish only if $G(x)$ contains positive and negative 
contributions, as in Eqn.(\ref{eq:g}).
This integral can vanish only if $G({\vec x})$ 
contains, at least, two length scales. 
At high temperatures, the length
\begin{eqnarray}
l_{1} \equiv |Re \{\beta\} |^{-1} \approx  \mu^{-1/2} (\mbox{ 
for $\mu \gg 2e$}). 
\end{eqnarray}
plays the role of the correlation length of the canonical short-range 
ferromagnet ({\it i.e.} 
%our action 
with no frustrating charge- $e=0$).  
Note that now, however, an additional correlation length appears: 
\begin{eqnarray}
l_{2} \equiv |Re \{\alpha\} |^{-1} \approx \frac{1}{\xi_{1} e}. 
%\gg \xi_{1}
\end{eqnarray}

For $\mu^{2} = 4e^{2}$, we find that $\alpha = \beta$, and the four 
poles merge in pairs.
When $\mu^{2}<4e^{2}$ the poles reside on a circle of radius $e^{1/4}$;
$Re\{\alpha^{2}\} = Re\{\beta^{2}\}= \frac{r}{2}$. This corresponds to
a pole at $k$ on the circle of radius $e^{1/2}$ at an angle $\theta$
with $\cos 2\theta
= - r/(2e)$. The analytic continuation for $\mu^{2}<4e^{2}$ is trivially
($\alpha \equiv \alpha_{1} +\imath \alpha_{2}= \beta^{*}$):
\begin{eqnarray}
 G({\vec  x}) = && \frac{k_{B}T}{4 \pi} 
\exp({-\alpha_{1}|{\vec  x}|}) \nonumber 
\\
 && \times \left[\frac{(\alpha_{2}^{2}-\alpha_{1}^{2})\sin \alpha_{2} 
|{\vec  x}|
      + 2 \alpha_{1} \alpha_{2} \cos
      \alpha_{2}|{\vec  x}|}{4\alpha_{1}\alpha_{2} |{\vec  x}|}\right]
\end{eqnarray}
where $\alpha \equiv \alpha_{1}+ \imath \alpha_{2}$. 
 
Putting all of the pieces together:
\begin{eqnarray}
   \mbox{ for } T > T_{1} \equiv k_{B}^{-1}(2 \pi)^{3}/[\int^{\Lambda} d^{3}k
    /({2e+\frac{e^{2}}{k^{2}}+k^{2})]} \nonumber
\\ > T_{c}(e=0) ; ~ \ell_{1} \neq \ell_{2}, 
\end{eqnarray}
\begin{eqnarray}
  {\mbox{ for } T < T_{1}} ~ ~  & \ell_{1} = \ell_{2}.
\end{eqnarray}
 
The temperature $T_{1}$, defined by $\mu(T=T_{1})=~2 e$, 
marks a dramatic crossover. 
At low temperatures ($ T < T_{1}$), the system possesses a
single correlation length $\xi = |\alpha_{1}|^{-1} =
2 {[\mu+ 2 e]}^{-1/2}$, and a single modulation length 
$L_{D} = 2 \pi / |\alpha_{2}| = 4 \pi {[-\mu+ 2e]}^{-1/2}$; 
at high temperatures 
($T>T_{1}$), the system possesses two distinct correlation lengths.  
When $T=T_{1}^{-}$, the modulation length diverges as 
$L_{D} \sim (T_{1}-T)^{-1/2}$. The crossover temperature
and the critical temperature read

\begin{eqnarray}
T_{1} \simeq 2 \pi^{2}[\Lambda-\frac{3 \pi}{4} e^{1/2}]^{-1}, \nonumber
\\ T_{c}(e=0) = \frac{2 \pi^{2}}{\Lambda},
\end{eqnarray}
with $\Lambda$ an upper momentum cutoff.

The mechanically oriented reader might note a simple analogy to the temporal
quadratic form for the damped oscillator, where two relaxation times appear 
for
$\gamma^{2} > 4m\kappa $ (with $\gamma$ , $\kappa$, and $m$ the dissipation
coefficient, the spring constant and the mass respectively).
At high temperatures, $\ell_{1} \approx \mu^{-1/2}$ plays 
the role of the usual
correlation length (for a free Landau-Ginzburg action with the
quadratic kernel $v(\vec{k}) = \mu+k^{2}$); this is the 
analogue of the damping time in an ``un-driven'' mechanical system.
At low temperatures ($ T < T_{1}$), the information previously encoded
in two correlation lengths at high temperatures is now manifest as a
single correlation length $\ell = |\alpha_{1}|^{-1} =
2\sqrt{\frac{1}{2 e + \mu}}$ and a single modulation length: $
L_{D} = 2 \pi / |\alpha_{2}| = 4 \pi \sqrt{ \frac{1}{2 e -
    \mu}}$.  The ``domain size'' $L_{D}$, the inverse characteristic
sinusoidal modulation length of the correlations, diverges at $T=T_{1}$, the
correlation length at $T_{c} = 0$.  The value registered by $\mu$ at
the avoided critical temperature $T_{c}(e=0)$ is roughly dimension
independent and pinned at $\mu \approx -e$; both $L_{D}$ and
$\ell$ are regular at $T_{c}(e=0)$. Note that, 
$lim _{e \rightarrow 0+} T_{1}(e) = T_{c}(e=0)$ 
(this statement can be well defined only in a slightly
 modified model such as $v(\vec{k}) = k^{2}+ e^{2}k^{-2} + \lambda 
\sum_{i \neq j} k_{i}^{2} k_{j}^{2}$ with small $\lambda$).  
In that limit $L_{D}$ makes an entrance
at the ``avoided critical point'' $T_{c}(e=0)$.
Such a crossover of a correlation length(s) into a
modulation length(s), or vice-versa, is a feature 
that is generically absent
in finite $Range$ interaction systems.

\subsection{Correlation Functions for Other Short
Range Frustrated Models and others}

Consider the interaction with the kernel
\begin{equation}
  v(\vec{k}) = [\Delta^{3}-\Delta_{0}^{3}]^{2},
\label{hex}
\end{equation}
where the lattice Laplacian
\begin{eqnarray}
\Delta(\vec{k}) = 2 \sum_{l=1}^{d} (1 - \cos k_{l}),
\end{eqnarray} 
with $\{k_{l}\}_{l=1}^{d}$ the momentum components.
As the lattice Laplacian
connects, in real space, two sites
that are a lattice distance apart,
the kernel of Eqn.(\ref{hex}) 
may link sites that are, at most,
six lattice constants apart.
It corresponds to a $Range=6$ interaction.
If the constant $\Delta_{0}>0$ then
upon expanding Eqn.(\ref{hex}), 
we will find both positive and 
negative contributions.
Expanding Eqn.(\ref{hex}),
we find both ferromagnetic
and anti-ferromagnetic interactions
that compete on different ranges.
When a chemical potential term is added 
to enforce normalization, we find that 
its minimum value is $\mu_{min} = \Delta_{0}^{6}$
(whence the mass gap vanishes).
For future reference let us denote the mass gap by
$c \equiv (\mu -\mu_{min})$. The poles of
the propagator
\begin{eqnarray}
  G(\vec{k}) = k_{B}T [c+v(\vec{k})]^{-1}, 
\end{eqnarray}
occur at
\begin{eqnarray}
\{  \Delta_{\alpha}^{\pm} \}_{\alpha=1}^{3} = [\Delta_{0}^{3} \pm \imath
  \sqrt{c}]^{1/3} \exp{(2 \pi \alpha \imath /3)}.
\end{eqnarray}

The poles $\{\Delta_{\alpha}^{\pm}\}$ form the vertices of a 
hexagon in the complex $k$ plane. Whenever $k_{i}^{3} = k_{o}^{3}$
with $k_{o}$ a given complex
number, 
then there must exist, at
least, two roots with different imaginary
values ($|Im\{k_{i}\}|$).  
The correlation lengths $\ell_{i} = |Im\{k_{i}\}|^{-1} $.
Two different correlation lengths $\ell_{1} \neq \ell_{2}$ 
can be identified at all $\mu \ge \mu_{min}$
($\mu=\mu_{min}$ only at $ T = T_{c} = 0 $). We note the existence of
multiple correlation lengths ${\ell_{i}}$ at all values of
$\Delta_{0}$ 
and temperatures (the mass gap $c$). Multiple correlations 
lengths are also present everywhere-
this also includes the unfrustrated system having $\Delta_{0}<0$! 
Multiple correlation lengths are 
generically present for
functions of $\Delta$ (or $k^{2}$ in the continuum) and are
accompanied by a discontinuity in the critical temperature $T_c$ 
in sufficiently high dimension whenever appropriate competition 
allows real roots $0 < \{\Delta_{i}\} < 4d $ when the mass gap
vanishes ($\mu =\mu_{min}$).
For the canonical $\Delta$ - polynomial, the system 
possesses several modulation lengths and several correlation 
lengths at all T; barring
few exceptions - their net number is conserved as temperature is
varied. Generically, all cross-over temperatures [$T_{i=1,2,...,p}$]
(including low dimensional systems which possess no critical behavior
for zero frustration), at which correlation lengths disappear and turn
into modulation lengths, tend continuously to the avoided critical
temperature (or its analytic continuation for low dimensions - in
high dimensions this temperature $\mu(T) = \mu^{*}$ becomes critical
for zero frustration).  For a general finite ranged $\Delta$ polynomial
kernel, the system possesses a fixed number of correlation lengths
and a fixed number of modulation lengths; there is no
sharp analogue of crossover temperatures
 $\{T_{i}\}$ wherein modulation lengths turn into
correlation lengths. These crossover temperatures are more prevalent 
for non-analytic functions of $\Delta$. The Yukawa Ferromagnet $v( \vec{k})=
\mu+(k^{2}+\lambda^{2})+\frac{e^{2}}{k^{2}+\lambda^{2}}$ (in real
space the last term gives rise to Yukawa like screened
interactions) has
$T_{c}(e=\lambda^{2})>0\mbox{ in $ d>4$ }$ and in any dimension
$T_{c}(e>\lambda^{2})=0$. Two correlation lengths appear for
$\mu^{2}>4e^{2}$ (including all unfrustrated ferromagnets ($e<0$)).

For the non-analytic kernel
 $v(\vec{k})= (|\vec{k}|-q)^{2}$,
\begin{eqnarray}
  \frac{1}{k_{B}T} \simeq \frac{1}{2 \pi^{2}}[ \Lambda + \frac{q^{2}-
    \mu}{\sqrt{\mu}}\{ \tan^{-1}(\frac{\Lambda-
    q}{\sqrt{\mu}})\nonumber
\\ +\tan^{-1}(q/\sqrt{\mu})\} +q \ln
  (\frac{\mu+(\Lambda-q)^{2}}{\mu+q^{2}})],
\end{eqnarray} 
with $\mu_{min} = 0$.

The correlator corresponding to $v(\vec{k})=
(|\vec{k}|-q)^{2}$, 
\begin{eqnarray}
G(\vec{x}) =\frac{1}{ 2 \sqrt{\mu}|\vec{x}|}  e^{-\sqrt{\mu}|\vec{x}|}
[q \sin qx + \sqrt{\mu} \cos qx] -\nonumber
\\ \frac{qI(|\vec{x}|)}{\pi^{2}|\vec{x}|},
\end{eqnarray}
where
\begin{eqnarray}
I(|\vec{x}|) = \int_{0}^{\infty} d \kappa \left [ \frac{ \kappa \exp[-\kappa |\vec{x}|]}{(\mu +q^{2}-\kappa^{2})^{2}+4q^{2}\kappa^{2}} \right ] >0.
\end{eqnarray}
Note the algebraic decay in this artificial non-analytic system.
Here the system displays avoided
critical behavior and exhibits one (two) modulation lengths  $2 \pi /q$ (and infinity),
which does not vary with temperature. 
Within the spherical model,
this behavior is an exception:
modulation lengths tend to vary with temperature.

With $t \equiv  (k_{B} T_{c}(q=0))^{-1}-(k_{B}T)^{-1})$, 
we find at low temperatures ($\mu^{3/2}$ terms neglected) for the 
$v(\vec{k}) = (|\vec{k}|-q)^{2}$ action,

\begin{eqnarray}
  t \approx (1+\frac{1}{2\pi^{2}}\frac{q}{\Lambda-q})\mu +
  \frac{1}{2\pi}\sqrt{\mu}\nonumber
\\ -\frac{1}{2\pi^{2}}q \ln
  (\frac{\mu+(\Lambda-q)^{2}}{\mu+q^{2}})-
  \frac{1}{2\pi}\frac{q^{2}}{\sqrt{\mu}}.
\end{eqnarray}

In the low temperature $\mu \rightarrow 0^{+}$ limit: $t \simeq
-\frac{1}{2\pi}\frac{q^{2}}{\sqrt{\mu}}$(as in any dimension $d \neq 3$
apart from a constant multiplicative factor). The large $n$ 
critical exponent $\nu_{q>0} =1$.  At
very low temperatures ($t<0$ with large $|t|$),
\begin{equation}
  \sqrt{r} \sim q^{2}/|t| \sim q^{2} \mbox{ } [\ell_{q=0}(-t)].
\end{equation}
If, by analogy to the standard quadratic part of Landau-Ginzburg actions,
we define a length $L_{D} \sim r^{-1/2}$, the inverse square root of
the mass gap which is the correlation
length in the canonical ($q=0$) case, then as $T \rightarrow
0^{+}, L_{D} \sim e^{-1} \ell_{q=0}^{-1}$ (with $q = e^{1/2}$).
We point here to the strong resemblance to the previously obtained result for the exact correlation
lengths appearing for the Coulomb frustrated 
ferromagnet with $v(\vec{k})= [e^{2}/k^{2}+k^{2}]$ at high
temperatures.

For the short-range (Teubner-Strey) correlator
\begin{eqnarray}
G^{-1}(\vec{k}) = a_{2}  +c_{1}k^{2}+c_{2}k^{4},
\end{eqnarray}
it is a simple matter to show that
\begin{eqnarray}
G(\vec{x}) \sim \frac{\sin \kappa x}{\kappa x} \exp[-x/\xi],
\end{eqnarray}
where
\begin{eqnarray}
\kappa = \sqrt{\frac{1}{2}\sqrt{\frac{a_{2}}{c_{2}}}- \frac{c_{1}}{4c_{2}}} \nonumber
\\ \xi^{-1} = \sqrt{\frac{1}{2}\sqrt{\frac{a_{2}}{c_{2}}}+ \frac{c_{1}}{4c_{2}}}.
\end{eqnarray}

In amphiphilic systems $a_{2}$ and $c_{1,2}$ are functions
of amphiphile concentration (as well as temperature).
In many simple
thermodynamical models,
short range interactions $k^{2}$
interactions thwarted by others,
the modulation length always 
varies with temperature:
in systems with long-range
interactions, the modulation 
length increases as temperature
is raised while the converse
generically holds for
short-range interactions.

When the kernel $v(\vec{k}) = k^{4}$, the real space correlator:
\begin{eqnarray}
G(\vec{x}) = -\frac{1}{4 \pi x \sqrt{\mu}} \exp{[-x/\xi]} \sin \kappa x.
\end{eqnarray}
Within the spherical model, this ferromagnetic system 
displays thermally induced oscillations.
At $T=0$ the (ferromagnetic)
ground state is unmodulated.

\section{Thermodynamic Functions in the vicinity 
of the disorder line}
\label{thermo_di}

\subsection{Non-Abelian Backgrounds}

For a global gauge field normalization, all multipliers
$\{\mu_{\ell}\}_{\ell=1}^{n} = const = \mu$.  The internal energy
density $U/N = n(k_{B}T - \mu)$.  AS before, the chemical potential 
$\mu$ may obtained by inverting $1/[k_{B}T] = \int \frac{d^{d}k}{(2 \pi)^{d}}
G(\vec{k})$. For fixed (non-dynamic) non-Abelian backgrounds, there
are no explicit high temperature crossovers- we find no
thermodynamic signatures of a disorder line.
The number of correlation lengths and
modulation lengths is temperature independent.
However, the long range physics is sinusoidal (governed 
by the poles $k_{\pm}$)
at low temperatures while it is nonsinuiodal (dominated
by the pole at $k_{0}$)
at intermediate temperature.

We now examine more complicated gauge theories of glasses.
To low orders in perturbation theory, 
our conclusions are not different. 
In the $l-$th representation of $SO(3)$ model,
the eigenvalues for $k \theta \gg 1$ are 
\begin{eqnarray}
\lambda_{m}(|\vec{k}|) \approx 
\theta^{2} [ (k \theta)^{2} + 2m (k \theta) \nonumber
\\ + \frac{1}{2} 
\{l(l+1) + m^{2} \}  \nonumber
\\ + \frac{1}{32[2(k \theta)+ m+1]}
[(l+m)(l-m+1)\nonumber
\\ \times (l+m-1) (l-m+2) \nonumber
\\ -(l-m)(l+m+1)\nonumber
\\ \times (l+m+2)(l-m+1)]].
\end{eqnarray}
To this order,  the equations 
\begin{eqnarray}
\mu + \lambda_{m}(|\vec{k}|) =0 
\end{eqnarray}
have a pair of complex conjugate
zeros and a single real zero
when $\mu > \mu_{\min}$. 
There is no crossover of a 
pure imaginary ($k$) poles into 
complex ones.
The computations are similar for an  
$SO(4)$ background gauge. 
When $ k \theta \ll 1$, the eigenvalues may be expanded
as polynomials in $(k \theta)$. The behavior of the 
correlation lengths and thermodynamic functions
is the same as that observed in 
the frustrated short range systems that we have 
discussed-- no merger of poles occurs, the 
number of correlation lengths and the number
of modulation lengths are temperature independent.
Thus fixed non-Abelian backgrounds do not lead
to explicit analytical (continuation) crossover in the 
thermodynamical functions.

\subsection{Thermodynamics near 
the Disorder Line for The Coulomb
Frustrated Ferromagnet}

In the spherical ($n \to \infty$) limit, the 
internal energy
\begin{eqnarray}
  \frac{U}{N} = \frac{1}{2N} \sum_{\vec{k}} [ \langle |S(\vec{k})|^{2}
\rangle
  v(\vec{k})\mbox{ }], \nonumber
\\ = \frac{1}{2N} \sum_{\vec{k}}
  \frac{k_{B}T}{v(\vec{k})+ \mu} v(\vec{k}) =\frac{1}{2}( k_{B}T - \mu).
\end{eqnarray}
The chemical potential, 
$\mu$, may be computed as a function of the temperature $T$ 
by inverting the spherical
equation of state $ \frac{1}{k_{B}T} = \frac{1}{nu}
\Sigma^{0}(r=\mu)$ with $\Sigma^{0}$ the zeroth order
self-energy.  $\Sigma^{0} (r=\mu) $ attains different
forms for $r^{2} > 4e^{2}$ (when $T>T_{1}$) and $r^{2} < 4e^{2}$ 
($T<T_{1}$). The relation 
$\tan^{-1}z = \frac{1}{2 i} \ln[ \frac{i-z}{i+z}]$
may be applied to relate the two natural 
functional forms. Although there are no divergences at the
merger point of the poles $T=T_{1}$, the explicit functional 
forms of the susceptibility $\langle
|S(\vec{q})|^{2} \rangle =
\frac{k_{B}T}{v(\vec{k})+\mu}$ and the specific heat $C_{V} =
\frac{\partial U}{\partial T}$ do change. For $T> T_{c}$, 
$\mu(T)$ is monotonically increasing in $T$.
\begin{equation}
  \frac{d \mu}{dT} =  \frac{1}{T^{2} \Pi(p=0)},
\end{equation}   
with the zero momentum polarization diagram $\Pi(p=0) = \frac{1}{(2
  \pi)^{3}} \int \frac{d^{3}k}{(v(\vec{k})+\mu)^{2}}$.  Here we see
that no specific heat extrema can be encountered at finite
temperatures in any
spherical model for $\mu > \mu_{\min}$ ($T>T_{c}$).  In $\lim n
\rightarrow \infty$ (and as shown, also to $O(1/n)$), $T_{1}$ is an analytic 
crossover
temperature. As many thermodynamic entities are integrals of pair 
correlators, the change in $G(\vec{x})$ above and below $T_{1}$ suggests
a crossover in certain thermodynamic quantities. 
We expect this crossover to persist to all orders in $1/n$.
We indeed find this to be the case to $O(1/n)$.
The underlying reason for this 
crossover is very transparent: In the factored propagators 
$1/[(k^{2}+\alpha^{2})(k^{2}+\beta^{2})]$, the poles have different 
explicit forms, if $T$ is above or below the disorder line temperature
$T_{1}$. The critical temperature
$T_{c}(e)$, as computed within the spherical model,
drops discontinuously. Above $T_{1}$ we have two correlation
lengths. Below $T_{1}$ there exists a 
single modulation length and a single correlation
length. As we have just outlined. even though
true long range order does not set it until
$T<T_{c}(e)$, an analytical crossover occurs
at $T=T_{1}(e)$. At low temperatures this crossover
temperature becomes the avoided critical
point $T_{c}(0)$. We have suggested that $T_{1}$ can be regarded
as an effective dimensional crossover temperature. 
At lower temperatures the behavior becomes
more and more like that of a one dimensional system.
At higher temperatures, the behavior is, in most
respects, similar to that of a three dimensional 
nearest neighbor ferromagnet.

\section{A generalized Mermin-Wagner Inequality}
\label{M-W-I}

\subsection{The Classical Case}

Our approach is the 
standard one.  We will 
keep it more general
instead of specializing
to anti/ferromagnetic order 
or to interactions 
of one special sort. With the 
notations introduced 
in Section(\ref{notations}),
we investigate $n$ component
spins on a lattice.
An applied magnetic field 
\begin{eqnarray}
\vec{h}(\vec{x}) = h \cos(\vec{q} \cdot \vec{x}) \hat{e}_{\alpha}
\end{eqnarray}
causes the spins
to take on their ground state values.

If $n=\alpha=2$ the unique spiral ground state ($\vec{S}^{g}(\vec{x})$), 
to which a low temperature  system would 
collapse to under the influence
of such a perturbation is
\begin{eqnarray}
S_{1}^{g}(\vec{x}) = \sin (\vec{q} \cdot \vec{x}).
\end{eqnarray}
When $n=3$ the ground state is not unique:
\begin{eqnarray}
S_{i<n}^{g}(\vec{x}) = r_{i} \sin(\vec{q} \cdot \vec{x}) \nonumber
\\ \sum_{i=1}^{n-1} r_{i}^{2}=1
\end{eqnarray}
and a magnetic field may be applied along two directions,
with all the ensuing steps trivially modified. 
With the magnetic field applied 
\begin{eqnarray}
H = \frac{1}{2} \sum_{\vec{x},\vec{y}} \sum_{i=1}^{n}
V(\vec{x}-\vec{y}) S_{i}(\vec{x}) S_{i}(\vec{y}) - 
\sum_{\vec{x}} h_{n}(\vec{x}) S_{n}(\vec{x}).
\label{last_H}
\end{eqnarray}
Note that the knowledge of the ground state
is not imperative in providing the forthcoming
proof \cite{smart}. We exploit the standard rotational
invariance of the measure
\begin{eqnarray}
\int d \mu~ ~\cdot    = ~ Z^{-1}  \int \prod_{\vec{x}} d^{n}S(\vec{x})
\delta(S^{2}(\vec{x})-1) e^{-\beta H}.
\end{eqnarray}
Note that this is not applicable 
to many other spin models (e.g. 
Dzyaloshinskii-Moriya interactions). The generators
 of rotation in the $[\alpha \beta]$ plane
are 
\begin{eqnarray}
L_{\alpha \beta} \equiv S_{\alpha} \frac{\partial}{\partial S^{\beta}}
-S_{\beta} \frac{\partial}{\partial S^{\alpha}}.
\end{eqnarray}
\begin{eqnarray}
0= \frac{d}{d \theta} \int d^{n}S ~\delta(\vec{S}^{2}-1)~
\nonumber
\\
f(S_{1},...,S_{\alpha} \cos \theta + S_{\beta} \sin \theta,...
\nonumber
\\, S_{\beta} \cos \theta - S_{\alpha} \sin \alpha,...,S_{n}).
\end{eqnarray}
\begin{eqnarray}
0= \int d^{n}S \delta (\vec{S}^{2}-1) L_{\alpha \beta} f(\vec{S}).
\end{eqnarray}
In the up and coming, $\perp$ will denote the
the projection 
along the $\beta$ direction.
We define the operators
\begin{eqnarray}
\vec{A}(\vec{k}) \equiv \sum_{\vec{x}} \exp[i \vec{k} \cdot \vec{x}]
\vec{S}_{\perp}(\vec{x}), \nonumber
\\ \vec{B}(\vec{k}) \equiv \sum_{\vec{x}} \exp[i (\vec{k} + \vec{q}) 
\cdot \vec{x}]
\vec{L}_{\vec{x}}(\beta H).
\end{eqnarray}
By the Schwarz inequality
\begin{eqnarray}
| \langle \sum_{i= \alpha, \beta} A_{i}^{*}B_{i} \rangle |^{2} ~\le ~~ \langle 
\sum_{i= \alpha, \beta} A_{i}^{*}A_{i} \rangle * 
\langle \sum_{i=\alpha,\beta} B_{i}^{*}B_{i} \rangle.
\end{eqnarray}
We will let $i=\alpha,\beta$ in the sum span a two element
subset of the $n$ spin components.
For any functional $C$:
\begin{eqnarray}
\vec{L}_{\vec{x}}(e^{-\beta H}C) = e^{-\beta H} \{ \vec{L}_{\vec{x}}(C)
+ C \vec{L}_{\vec{x}}(-\beta H)\}.
\end{eqnarray}
\begin{eqnarray}
0 = \int \prod_{\vec{x}}~ d^{n}S(\vec{x})~ \delta (\vec{S}^{2}(\vec{x})-1)
~\vec{L}_{\vec{x}}[e^{-\beta H}C]
\end{eqnarray}

\begin{eqnarray}
\langle C B(\vec{p}) \rangle = \langle \sum_{\vec{x}} \exp[i \vec{p}
\cdot \vec{x}] ~\vec{L}_{\vec{x}}(C) \rangle
\end{eqnarray}

\begin{eqnarray}
\sum_{i= \alpha,\beta}  \langle L_{\vec{x}}^{i}(L_{\vec{y}}^{i}(\beta H)) 
\rangle~ =
 \beta \langle [\sum_{i=\alpha,\beta} S_{i}(\vec{x}) S_{i}(\vec{y})
 V(\vec{x}-\vec{y})
\nonumber
\\ - h(\vec{x}) S_{n}(\vec{x})] \rangle
\end{eqnarray}
\begin{eqnarray}
\langle \vec{B}(\vec{k})^{*} \cdot \vec{B}(\vec{k}) \rangle 
= \beta \sum_{\vec{x},\vec{y}} 
 \{ (\cos (\vec{k} + \vec{q}) \cdot (\vec{x}-\vec{y})-1) \nonumber
\\ \Big[ \sum_{i=\alpha, \beta} \langle S_{i}(\vec{x}) S_{i}(\vec{y})
\rangle  \Big] 
 V(\vec{x}-\vec{y}) \}   \nonumber
\\ -
h(\vec{x}) \langle S_{n}(\vec{x}) \rangle 
 ~~\ge~ 0
\end{eqnarray}

Henceforth, for simplicity, we specialize 
to $n=2$. Fourier expanding the interaction
kernel 
\begin{eqnarray}
V(\vec{x}-\vec{y}) = \frac{1}{N} \sum_{\vec{t}} v(\vec{t}) e^{i \vec{t} \cdot
(\vec{x}-\vec{y})},
\end{eqnarray} 
and substituting
\begin{eqnarray}
\langle \vec{S}(\vec{x}) \cdot \vec{S}(\vec{y}) \rangle = \frac{1}{N^{2}}
\sum_{\vec{u}} \langle |\vec{S}(\vec{u})|^{2} \rangle e^{i \vec{u} \cdot 
(\vec{x} - \vec{y}) },
\end{eqnarray}
we obtain 
\begin{eqnarray}
0 \le \langle \vec{B}(\vec{k})^{*} \cdot  \vec{B}(\vec{k}) \rangle
\equiv  
\beta \Delta_{\vec{k}}^{(2)} E -\beta h_{\vec{q}} \langle
S_{n}(-\vec{q}) \rangle = \nonumber
\\ \frac{\beta}{2N} \sum_{\vec{u}} 
 \Big[ v(\vec{u} + \vec{k}) + v(\vec{u}
- \vec{k}) \nonumber
\\ - 2 v(\vec{u}) \Big] \langle |\vec{S}(\vec{u})|^{2} \rangle
- \beta h_{\vec{q}} \langle S_{n}(-\vec{q}) \rangle 
\label{B*B}
\end{eqnarray}
where $\Delta_{\vec{k}}^{(2)} E$ measures 
the finite difference 
of the internal energy with respect to 
a boost of momentum $\vec{k}$. 
\begin{eqnarray}
\langle \vec{A}(\vec{k})^{*} \cdot \vec{A}(\vec{k}) \rangle = 
\sum_{\vec{x},\vec{y}}
\langle \vec{S}_{\perp}(\vec{x}) \cdot \vec{S}_{\perp}(\vec{y})
\rangle
\nonumber
\\ \times \exp[i \vec{k} \cdot (\vec{x}-\vec{y})].
\end{eqnarray}
\begin{eqnarray}
 \langle \vec{A}(\vec{k})^{*} \cdot \vec{B}(\vec{k}) \rangle= 
\langle \sum_{i,\vec{x}}
L_{\vec{x}}^{i}(\vec{S}_{\perp}(\vec{x})) \exp[i (\vec{k} + \vec{q}) 
\cdot \vec{x}] \rangle
= m_{\vec{q}},
\nonumber
\end{eqnarray}
where
$m_{\vec{q}} \equiv \langle S_{n}(\vec{q}) \rangle$
and, as noted earlier, $\perp$ refers to the $i=1$ spin direction 
orthogonal to $i=n=2$.
Note that with our convention
for the Fourier transformations,
a macroscopically modulated state of
wave-vector $\vec{q}$, the magnetization
$m_{q} ={\cal{O}}(N)$
as is the energy difference in Eqn.(\ref{B*B}).
Trivially rewriting the 
Schwarz inequality and
summing over all
momenta $\vec{k}$,
\begin{eqnarray}
\sum_{\vec{k}} \frac{
\langle \vec{A}(\vec{k})^{*} \cdot \vec{B}(\vec{k})
\rangle}{\langle 
\vec{B}(\vec{k})^{*} \cdot \vec{B}(\vec{k}) \rangle}
\le \sum_{\vec{k}} \langle \vec{A}(\vec{k})^{*} \cdot \vec{A}(\vec{k})
\end{eqnarray}
which explicitly reads
\begin{eqnarray}
2 N |m_{\vec{q}}|^{2}  \Big( \beta  \sum_{\vec{k}} 
 ( \langle |\vec{S}(\vec{u})|^{2} \rangle  [v(\vec{k}+\vec{u}) \nonumber 
\\ +v(\vec{u}- \vec{k})-2 v(\vec{u})]+ 2 |h||m_{q}| 
\Big)^{-1}\nonumber
\\ \le N \sum_{\vec{k}} \sum_{\vec{x},\vec{y}} \langle \vec{S}_{\perp}(\vec{x})
\cdot \vec{S}_{\perp}(\vec{y}) \rangle e^{i \vec{k} \cdot (\vec{x}-\vec{y})}
\nonumber 
\\ = N \sum_{\vec{x}} \langle \vec{S}_{\perp}^{2}(\vec{x}) \rangle.
\label{ineq}
\end{eqnarray}

Explicitly, as the integral $\int^{|\vec{k}| > \delta}$ 
$\frac{d^{d}k}{(2 \pi)^{d}} ...$ is non-negative 
(as $\langle \vec{B}(\vec{k})^{*} \cdot \vec{B}(\vec{k}) \rangle \ge 0$
the denominator in Eqn.(\ref{ineq})
is positive for each individual 
value of $\vec{k}$), and
as $\langle \vec{S}_{\perp}^{2}(\vec{x}) \rangle  \le 1$,
we obtain in the thermodynamic limit
\begin{eqnarray}
\frac{2}{\beta} |m_{\vec{q}}|^{2}  \int^{|\vec{k}| < \delta}
\frac{d^{d}k}{(2 \pi)^{d}} \Big[ \int \frac{d^{d}u}{(2 \pi)^{d}} \nonumber
\\  \langle |\vec{S}(\vec{u})|^{2} \rangle  (v(\vec{k}+\vec{u}) 
+v(\vec{k}-\vec{u})-2 v(\vec{u}))  \nonumber
\\ +2 |h| |m_{q}| \Big]^{-1}
 \le 1.
\label{main}
\end{eqnarray}
Taking $\delta$ to be small we may bound from above 
(for each value of $\vec{k}$) the positive
denominator in the square brackets 
and consequently
\begin{eqnarray}
 \int^{|\vec{k}| < \delta}
\frac{d^{d}k}{(2 \pi)^{d}} \Big[ \int \frac{d^{d}u}{(2 \pi)^{d}} \nonumber
\\  \langle |\vec{S}(\vec{u})|^{2} \rangle  (v(\vec{k}+\vec{u}) 
+v(\vec{k}-\vec{u})-2 v(\vec{u}))  \nonumber
\\ +2 |h| |m_{q}| \Big]^{-1}
\ge \int^{|\vec{k}| < \delta}
\frac{d^{d}k}{(2 \pi)^{d}} \Big[ \int \frac{d^{d}u}{(2 \pi)^{d}} \nonumber
\\ A_{1}  k^{2} \lambda_{\vec{u}} \langle |\vec{S}(\vec{u})|^{2} \rangle  
+2 |h| |m_{q}| \Big]^{-1},
\end{eqnarray}
with $\lambda_{\vec{u}}$ chosen to be the largest 
principal eigenvalue of the $d \times d$ 
matrix $\partial_{i} \partial_{j} [v(\vec{u})]$,
and $A_{1}$ a constant. 
For a twice differentiable $v(\vec{u})$, and for $|\vec{k}| \le \delta$ 
where $\delta$ is finite,
\begin{eqnarray}
(v(\vec{k}+\vec{u}) 
+v(\vec{k}-\vec{u})-2 v(\vec{u})) 
\le A_{1} \lambda_{\vec{u}} k^{2} \le B_{1} k^{2}
\end{eqnarray}
for all $\vec{u}$ within the Brillouin Zone  
with the additional positive constant $B_{1}$
introduced \cite{precision}. Here we
reiterate that $\langle \vec{B}^{*}(\vec{k}) \cdot \vec{B}(\vec{k}) \rangle$
of Eqn.(\ref{B*B}) is positive definite
and consequently the
bound derived is 
powerful.

In $d \le 2$, the integral
$\int^{|\vec{k}| < \delta} \frac{d^{d}k}{B_{1}k^{2}}$
diverges making it possible to satisfy  
eqn.(\ref{main}), at finite temperatures,
when the external magnetic
field $h \rightarrow 0$ only if 
the magnetization $m_{q}=0$.
If finite size effects 
are restored, in a system
of size $N= L \times L$ where 
the infrared cutoff in
the integral is ${\cal{O}} (\frac{2 \pi}{L}, \frac{2 \pi}{L})$
the latter integral diverges as ${\cal{O}}(\ln N)$.
This implies that the upper bound on $|m_{\vec{q}}|$
scales as ${\cal{O}}(N / \sqrt{\ln N})$, much lower than
the $ {\cal{O}}(N)$ requisite for finite 
on-site magnetization. For further 
details see \cite{smart}. 
If there are $M \ge 2$ pairs
of minimizing modes and $2p +1 \ge  n \ge 2p$
(with an integer $p$) then we may apply an infinitesimal 
symmetry breaking magnetic field along, at most, $\min\{p,M\}$ 
independent spin directions ($\alpha$). Employing the
spin rotational invariance within each  plane 
 $[\alpha \beta]$ associated with any individual 
mode $\vec{\ell}$ (au lieu of a specific $\vec{q}$)
we may produce a bound similar that 
in Eqn.(\ref{main}) wherein $\langle |\vec{S}(\vec{u})|^{2}
\rangle$ will be replaced by $\sum_{i= \alpha,\beta} 
\langle |S_{i}(\vec{u})|^{2} \rangle$ and $|m_{\vec{q}}| \rightarrow
|m_{\alpha}(\vec{\ell})|$.

\subsection{The Quantum Case}

The finite temperature behavior 
of a quantum system is, in many 
respects, similar to that of a 
classical system. The quantum system
is also invariant under rotations 
with $\hat{S}^{2}(\vec{x}) = S(S+1)$.

Alternatively, one could directly tackle the
$n=3$ quantum case by applying the Bogoliubov
inequality
\begin{eqnarray}
\frac{\beta}{2} \langle \{A,A^{\dagger}\} \rangle * \langle
[~[C,H],C^{\dagger}]\rangle \ge | \langle [C, A]\rangle |^{2}
\end{eqnarray}

with $[~,~]$ and $\{~,~\}$ the commutator and anticommutator
respectively. From this inequality,
it follows that $ \langle
[~[C,H],C^{\dagger}]\rangle$
is positive definite. 
In particular, for any six
operators $\{A_{1},A_{2},A_{3},C_{1},C_{2},C_{3}\}$,
\begin{eqnarray}
\frac{\beta}{2} (\sum_{a=1}^{3} 
 \langle \{A_{a},A_{a}^{\dagger}\} \rangle)
(\sum_{a} \langle
[~[C_{a},H],C^{\dagger}_{a}]\rangle)
\ge  \nonumber
\\ \sum_{a} | \langle [C_{i}, A_{i}]\rangle |^{2}
\label{Bog_symm}
\end{eqnarray}

Setting $A_{1}= S_{2}(\vec{q}-\vec{k})$ and 
$C_{1}= S_{1}(\vec{k})$
we will once
again obtain Eqn(\ref{main}) 
with the classical spins replaced 
by their quantum counterparts.

Rather explicitly, employing  
\begin{eqnarray}
[S^{\alpha}(\vec{k}), S^{\beta}(\vec{k}^{\prime})] = i 
\epsilon^{\alpha \beta \gamma} S_{\gamma}(\vec{k}+ \vec{k}^{\prime}),
\end{eqnarray}
we find for the Hamiltonian
of Eqn.(\ref{last_H}) (with $n=3$),
\begin{eqnarray}
[[C_{1},H],C_{1}^{\dagger}] = 
\frac{1}{2N} \sum_{\vec{k}^{\prime}} 
(S_{2}(\vec{k}^{\prime}) S_{2}(-\vec{k}^{\prime})
+ S_{3}(\vec{k}^{\prime}) S_{3}(-\vec{k}^{\prime})) \nonumber
\\ \times
[v(\vec{k}+ \vec{k}^{\prime}) - 2 v(\vec{k}) 
+ v(\vec{k}^{\prime} - \vec{k})] \nonumber
\\ 
+ \frac{1}{N} \sum_{\vec{k}^{\prime}} h_{3}(\vec{k}^{\prime}) 
S_{3}(-\vec{k}^{\prime}). 
\nonumber
\end{eqnarray}

Similarly,
\begin{eqnarray}
|\langle [C_{1}, A_{1} ] \rangle|^{2} = |\langle S_{3}(\vec{q}) \rangle|^{2},
\end{eqnarray}
the squared magnetization along the z (or 3) direction for a
mode $\vec{q}$, 
and
\begin{eqnarray}
\sum_{\vec{k}} \langle \{ A_{1}, A^{\dagger}_{1} \} \rangle
= 2 \sum_{\vec{k}} \langle S_{2}(\vec{k}) S_{2}(-\vec{k}) \rangle \nonumber
\\  = 2 \langle [S_{2}(\vec{x}=0)]^{2} \rangle.
\end{eqnarray}

Next, let us cyclically
set, $A_{2} = S_{3}(\vec{q}-\vec{k}), 
C_{2} = S_{2}(\vec{k}), A_{3} = S_{1}(\vec{q} - \vec{k})$, 
and $C_{3} = S_{3}(\vec{k})$.
The commutators associated
with these operators
are all identically the same
apart from a uniform 
cyclic permutation
of all spin components
involved.

Trivially rewriting 
the symmetrized Bogoliubov inequality 
Eqn.(\ref{Bog_symm})
and summing over all
modes $\vec{k}$,
\begin{eqnarray}
\frac{\beta}{2} \sum_{\vec{k},a} 
\langle \{A_{a}, A^{\dagger}_{a} \} \rangle
\ge \sum_{\vec{k}} \frac{  \sum_{a}| \langle  
[C_{a}, A_{a}] \rangle |^{2}}{ \sum_{a} \langle [[C_{a},H],
C^{\dagger}_{a}] \rangle}.
\end{eqnarray}

Replacing the $\vec{k}$ sums by
integrals in the thermodynamic
limit, and
employing 
the positivity
of $\langle [[C_{a},H],
C_{a}^{\dagger}] \rangle$
that follows from the
Bogoliubov inequality
for each individual
value of $\vec{k}$, we find the trivial
quantum analogue of
Eqn.(\ref{main}),
\begin{eqnarray}
\frac{1}{2 \beta} |m_{\vec{q}}|^{2}  \int^{|\vec{k}| < \delta}
\frac{d^{d}k}{(2 \pi)^{d}} \Big[ \int \frac{d^{d}u}{(2 \pi)^{d}} \nonumber
\\  \langle |\vec{S}(\vec{u})|^{2} \rangle  (v(\vec{k}+\vec{u}) 
+v(\vec{k}-\vec{u})-2 v(\vec{u}))  \nonumber
\\ +|h| |m_{q}| \Big]^{-1}
 \le S(S+1).
\label{main_q}
\end{eqnarray}
Apart from the simple scaling
factor of $4S(S+1) = 2 \sum_{a} 
\langle \{A_{a}, A^{\dagger}_{a} \} \rangle$
by comparison to the classical case,
there is no difference
between this inequality
and its classical counterpart
in the zero field limit.
We symmetrized the Bogoliubov inequality
(Eqn.(\ref{Bog_symm})) in order
to avoid the appearance
of only two transverse 
spin components in the
Fourier weights $\langle |S_{i}(\vec{q})|^{2} 
\rangle$ so as to give 
the resulting Mermin-Wagner inequality
a transparent physical meaning
associated with the energy 
boost differences $\Delta_{\vec{k}}^{(2)} E$
which are symmetric in all
spin indices. From here onward the discussion 
can proceed as in the classical
case.

\section{Mermin-Wagner Bounds in High Dimensions}
\label{M-W-I-boost}

In any dimension,  
Eqn.(\ref{main}) reads in the limit $h \to  0$ 
\begin{eqnarray}
2 |m_{\vec{q}}|^{2} T \int \frac{d^{d}k}{(2 \pi)^{d}} ~ ~ 
\frac{1}{\Delta_{\vec{k}}^{(2)}E} \le 1
\label{general**}
\end{eqnarray}
with the shorthand defined by
Eqn.(\ref{B*B}).
By parity invariance and noting that   
$\{\vec{S}(\vec{x})\}$ are real,
\begin{eqnarray}
\sum_{\vec{u}} v(\vec{k} + \vec{u}) \langle
|\vec{S}(\vec{u})|^{2} \rangle = \sum_{\vec{u}} v(\vec{k} - \vec{u})
\langle
|\vec{S}(\vec{u})|^{2} \rangle.
\end{eqnarray}
Thus the denominator of Eqn.(\ref{general**}) 
reads $\Delta_{\vec{k}}^{(2)}E = [2(E_{\vec{k}}-E_{0})]$
where $E_{\vec{k}}$ is the internal 
energy of system after undergoing a 
boost of momentum $\vec{k}$ and $E_{0}$ 
denotes the internal energy of the 
un-boosted system.
In some instances when the dispersion relation 
about an assumed zero temperature ground
state is inserted into Eqn.(\ref{general**}),
we will find that the integral in Eqn.(\ref{general**})
diverges: At arbitrarily
low temperatures we cannot assume the zero temperature
ground state with the natural dispersion relation
$\Delta E$ for fluctuations about it.
A case in point is the dispersion relation
for the Coulomb
Frustrated Ferromagnet. The denominator
in Eqn.(\ref{main}) is a finite temperature 
extension of the $T=0$ dispersion  
relations ($\Delta_{\vec{k}}^{(2)} E$).
In general, at zero temperature,
\begin{eqnarray}
\langle |\vec{S}(\vec{k})|^{2} \rangle = \frac{N^{2}}{2}
[\delta_{\vec{k},\vec{q}}+ \delta_{\vec{k},-\vec{q}}]
\label{ideal-occ}
\end{eqnarray}
and the integral 
in Eqn.(\ref{general**})
becomes Eqn.(\ref{MY_IN}).

Whenever Eqn.(\ref{MY_IN}) diverges, an assumption
of an almost ordered ground state at arbitrarily
low temperatures ($T = 0^{+}$) 
with the ensuing zero temperature 
dispersion relation 
($\Delta_{\vec{k}}^{(2)}E = [2(E_{\vec{k}}-E_{0})]$) 
about that ordered state is flawed: Eqn.(\ref{general**}) 
is strongly violated.
This poses no problem for most 
canonical $ d>2$ dimensional models
where the integral in Eqn.(\ref{MY_IN})
is finite. For the three dimensional Coulomb 
Frustrated Ferromagnet and other high dimensional 
models the divergence of this integral
hints possible non-trivialities.
The 
average decoherence (or
relaxation) time 
scale $\langle \tau \rangle $ 
may diverge if we assume a 
zero temperature 
dispersion of fluctuations
about an ideal crystal.
A divergent decoherence time scale suggests the 
absence of broken translational
symmetry. Formally,
\begin{eqnarray}
\int \frac{d^{3}k}{(2 \pi)^{3}} \tau_{k} = \tau(\vec{r}=0).
\end{eqnarray} 
In order
for magnetization $m_{q} = {\cal{O}}(N)$
to arise, the average of the inverse boost energy
over all of $\vec{k}$ space
\begin{eqnarray}
\tau \equiv \langle \frac{1}{\Delta_{\vec{k}}^{(2)}E} 
\rangle_{\vec{k}} \le  {\cal{O}}(T^{-1}),
\label{last-ate}
\end{eqnarray}
at all $0<T<T_{c}$.
In most $d >2$ systems
this is trivially
satisfied with the average
bounded by a constant at zero temperature.
This average 
diverges at $T=0$ whenever Eqn.(\ref{MY_IN})
does. There are two possibilities:

(i) The system is ordered at all temperatures $T<T_{c}$
in which case, the thermodynamic average of Eqn.(\ref{last-ate}) 
is finite for all $T>0$ and $\tau$ is non-analytic 
at $T=0$.

(ii) The system is disordered at all finite temperatures
and orders classically only at $T=0$.

The first possibility ((i)) was argued for
by a non-rigorous yet  
elegant diagrammatic analysis
by Brazovskii \cite{Brazovskii} 
long ago: Thermal fluctuations,
on their own, may enhance (or generate) cubic terms fortifying
(or triggering weak) first order transitions.
This cannot be ruled out by
the rigorous Mermin-Wagner inequalities
that we derived.
As reiternated, if the fluctuation integral
diverges then we may not obtain the 
low temperature $T=0^{+}$ 
dispersion by assuming a 
nearly perfectly ordered 
state. This does not
rigorously preclude order.
Order, if it exists at
arbitrarily low temperatures
must display non-trivial 
excitation spectra about it. 
If such a 
possibility arises,
it might be immaterial
if the system is permanently
frozen into a glass before
reaching an equilibrium 
thermodynamic transition.
In both cases, $\tau(T)$ is
non-analytic at $T=0$.
The integral of Eqs.(\ref{MY_IN},\ref{last-ate}) 
has a suggestive physical interpretation. If 
the quantum spin system is subjected to a boost of momentum $\vec{k}$, then 
\begin{eqnarray}
1/\Delta_{\vec{k}}^{(2)}E \equiv \tau_{\vec{k}}
\end{eqnarray}
is the characteristic lifetime of the
excited state. The average in Eqn.(\ref{last-ate}) is 
the characteristic relaxation (or decoherence) time 
of the system
averaged over magnons of all possible 
momenta. Whenever the average characteristic relaxation
time 
\begin{eqnarray}
\langle \tau_{\vec{k}} \rangle = \int
\frac{d^{d}k}{(2 \pi)^{d}} \Big[ \int \frac{d^{d}u}{(2 \pi)^{d}} \nonumber
\\  \langle |\vec{S}(\vec{u})|^{2} \rangle  (v(\vec{k}+\vec{u}) 
+v(\vec{k}-\vec{u})-2 v(\vec{u})) \Big]^{-1}
\end{eqnarray}
diverges then by our generalized 
inequality 
\begin{eqnarray}
2|m_{q}|^{2}T \langle \tau_{\vec{k}} \rangle = 2|m_{q}|^{2}T  
\tau_{r=0} \le 1
\end{eqnarray}
the system does not order
in such a way that the 
fluctuation dispersion
about any viable ground
state is valid. Moreover, 
as we show in the text the system displays
divergent glassy dynamics. 
The characteristic
divergent relaxation times in 
glassy systems suggest
a divergent decoherence
time and an inability
to write a wave function
of a quantum glass.  
The glassy state
may be specified 
by a density matrix \cite{obs-jorg}.

$^{*}$ Present address: Theoretical Division, Los Alamos National Laboratory, 
Los Alamos, NM 87545


\begin{thebibliography}{1}
\bibitem{the_wave} In a similar spirit, if we demand 
the matching, upon rolling, 
of ideal wave-functions present on 
spheres, the action would be the same
with $\rho$ replaced by a ``wave-function'' 
\begin{equation}
  \Psi(\vec{x},\hat{n}) \equiv \sum_{\ell,m} Q_{\ell,m}(\vec{x})
  Y_{\ell}^{m}(\hat{n}).
\label{wave-1}
\end{equation}
The global normalization reads:
\begin{equation}
  1= \int d^{2}x \sum_{\ell,m} |Q_{\ell, m} (\vec{x})|^{2},
\label{wave-2}
\end{equation}
which is, en effect, the spherical constraint of the large $n$
spin models which we will come back to later
on. The Hamiltonian will be the 
same with the kinetic energy term $(-\nabla^{2} \Psi)$
replaced a squared covariant derivative $(-D^{2} \Psi)$.
All stated
can be applied to such problems.
\bibitem{smart}
Note that even if we had not known the ground states
we could still prove that there is
no magnetization. All we need to do is to 
apply an infinitesimal magnetic
field 
\begin{equation}
\vec{h}(\vec{x}) = h \vec{S}^{ground-state}(\vec{x})
\end{equation}
with as yet an unknown ground state $\vec{S}^{ground-state}(\vec{x})$ .
Replacing any appearance of the minimizing 
wave-vector $\vec{q}$ in the Schwarz inequality 
and the definition of $B(\vec{k})$ by the more general 
wave-vector $\vec{\ell}$ and setting
$h \to 0^{+}$ one would arrive at 
the conclusion that $m_{\vec{\ell}}=0$
for each mode $\vec{\ell}$. 
Consequently, 
\begin{eqnarray}
\langle S_{n}(\vec{x}) \rangle = \frac{1}{N} \sum_{\vec{\ell}} e^{i
\vec{l} \cdot \vec{x}} m_{\ell} = 0
\end{eqnarray}


Stated alternatively, by Parseval's theorem,
\begin{eqnarray}
\sum_{\vec{x}} \langle \vec{S}(\vec{x}) \rangle^{2} = \frac{1}{N^{2}} 
\sum_{\vec{l}} \langle \vec{S}(\vec{l}) \rangle \langle 
\vec{S}(-\vec{l}) \rangle. 
\label{Pars}
\end{eqnarray}

Taking careful note of the system size ($N$) 
in the fluctuation integrals and the $N$ terms 
in the summand of Eqn.(\ref{Pars}),
\begin{eqnarray}
\sum_{\vec{x}} \langle \vec{S}(\vec{x}) \rangle^{2} \le {\cal{O}}
(\frac{N}{\ln N})
\end{eqnarray}
and  $|\langle \vec{S}(\vec{x}) \rangle|$
at any site $\vec{x}$ diminishes as
${\cal{O}}(\sqrt{1/\ln N})$,
vanishing in the thermodynamic ($ N \rightarrow \infty$)
limit.
\bibitem{precision}
More precisely, everything 
to be claimed holds for 
the more general class of functions
$v(\vec{k})$ for which $\max_{\vec{u} \in B.Z.} [v(\vec{k}+ \vec{u}) 
+ v(\vec{k} - \vec{u}) - 2
v(\vec{u})] \le B_{1} k^{2}$
is obeyed for 
all momenta $\vec{k}$
with some finite positive constant 
$B_{1}$.
\end{thebibliography}
\end{document}